\documentclass[a4paper,11pt]{article}
\pdfoutput=1 

\usepackage{jcappub} 

\usepackage[T1]{fontenc} 
\usepackage{multirow}

\usepackage[colorinlistoftodos]{todonotes}
\usepackage{soul}
\usepackage{hyperref}
\setstcolor{blue}
\graphicspath{ {./img/} }
\allowdisplaybreaks
\title{\boldmath CTA Sensitivity on TeV scale \\ Dark Matter Models \\ with Complementary Limits \\ from Direct Detection }

\author[a*]{C. Duangchan,\note[*]{Corresponding author.}}
\author[b*]{C. Pongkitivanichkul,}
\author[c*]{P. Uttayarat,}
\author[d,e*]{A. Jardin-Blicq,}
\author[a]{M. Wechakama,}
\author[b,e]{T. Klangburam,}
\author[c]{W. Treesukrat,}
\author[b]{D. Samart,}
\author[e]{U. Sawangwit,}
\author[f,g]{A. Aguirre-Santaella,}
\author[f,g]{and M. A. S\'{a}nchez-Conde}

\affiliation[a]{Kasetsart University, 50 Ngamwongwan Rd., Lat Yao, Chatuchak, Bangkok 10900, Thailand}
\affiliation[b]{Khon Kaen University, 123 Mitraphap Rd., Khon Kaen, 40002, Thailand}
\affiliation[c]{Srinakharinwirot University, 124 Sukhumvit 23 Rd., Wattana, Bangkok, 10110, Thailand}
\affiliation[d]{Chulalongkorn University, 254 Phayathai Rd., Pathumwan, Bangkok 10330, Thailand}
\affiliation[e]{National Astronomical Research Institute of Thailand, Don Kaeo, MaeRim, Chiang Mai 50180, Thailand}
\affiliation[f]{Instituto de F\'{i}sica Te\'{o}rica UAM-CSIC, Universidad Aut\'{o}noma de Madrid, C/ Nicol\'{a}s Cabrera, 13-15, 28049 Madrid, Spain}
\affiliation[g]{Departamento de F\'{i}sica Te\'{o}rica, M-15, Universidad Aut\'{o}noma de Madrid, E-28049 Madrid, Spain}

\emailAdd{chaimongkol.d@ku.th}
\emailAdd{chakpo@kku.ac.th}
\emailAdd{patipan@g.swu.ac.th}
\emailAdd{armelle@narit.or.th}

\abstract{With ever increasing pressure from collider physics and direct detection experiments, particle physics models of TeV scale dark matter are gaining more attention. In this work, we consider two realizations of the class of scalar portal dark matter scenarios -- the inverse seesaw model and the inert doublet model. Observations by the Cherenkov Telescope Array (CTA) of very-high-energy $\gamma$ rays from dark matter annihilation in the context of these models are simulated for the Draco and Sculptor dwarf spheroidal galaxies, and later analyzed using \texttt{ctools}. We study the potential of CTA for the 5$\sigma$ detection of a dark matter annihilation signal. 
In the absence of a signal, we also derive the 2$\sigma$ upper limits on the annihilation cross-section. We compare our projected CTA sensitivity against the projected sensitivity of the next generation of direct detection experiment, i.e. XENONnT. Although the limits from CTA are significantly improved compared with the previous generations of $\gamma$-ray experiments, they are still $\sim2$ orders of magnitude above the thermal relic cross-section for the considered targets. In the case of the inverse seesaw model, the constraint from the future direct detection experiment XENONnT is much weaker than the CTA sensitivity, whereas for the inert doublet model, XENONnT gives a bound an order of magnitude stronger compared to the CTA limits.}

\begin{document}
\maketitle
\flushbottom

\section{Introduction}

A recent analysis of the cosmic microwave background indicates that baryonic matter only makes up 4.9\% of the $\Lambda$CDM universe. The rest is dominated by the dark components, 26.8\%  dark matter (DM) and 68.3\% dark energy~\cite{Planck_Cosmological_parameters}.  Although a lot of indicative evidence has been accumulated such as rotation curves of galaxies, gravitational lensing and structure formation, the mystery of DM still remains for more than eight decades since Frizt Zwicky first coined the term (see e.g.~\cite{Bertone_etal_Dark_Matter_Review} for a review). Despite this, DM has managed to escape all forms of detection and although a lot still remains unclear, some advancements have been made in understanding the unknown that these dark components represent~\cite{Jungman1996SupersymmetricDM, Bertone2005ParticleDM, Feng2010DMCandidates}. One such advancement is the definitive conclusion that DM lies beyond the Standard Model (SM) of particle physics. The most prominent candidate is the weakly interacting massive particle or ``WIMP''. The typical annihilation cross-section of the WIMPs turns out to be around the same scale as the thermal relic cross-section $\langle\sigma v\rangle\simeq2.6\times10^{-26}$ cm$^3$/s, which is known as ``the WIMP miracle''~\cite{Arcadi2018WIMPs}.

The lack of detection is not due to a lack of trying. Many different forms of experiments have been conducted with the aim of shedding light on DM~\cite{Bertone2018DMNewEra}. Some are anticipating DM particles to collide with atomic nuclei, generically referred to as a direct detection experiment, resulting in recoil energy that might give some indication about the nature of DM.  Some others are attempting to create DM particles inside a particle accelerator. While these types of experiments have failed to produce clues about what DM is, analyses of a wide range of available data~\cite{EuropeanStrategyforParticlePhysicsPreparatoryGroup:2019qin} have established a deeper understanding regarding what DM is not. 
Data from both direct detection experiments and collider searches have produced relevant constraints that help to frame our understanding of DM, which is also true for research on indirect detection.  A core assumption about indirect detection experiments is that DM can annihilate, or decay, into SM particles such as photons, leptons, quarks and neutrinos. Based on this, experiments such as PAMELA, AMS02, and IceCube have reported constraints of DM masses ranging from MeV to GeV~\cite{PAMELA:2017bna,AMS:2021nhj,Baur:2019jwm}. Gamma-ray experiments such as the Fermi Large Area Telescope (Fermi-LAT), as well as Imaging Atmospheric Cherenkov Telescopes (IACTs) have also provided constraints of DM properties with masses ranging up to few TeV~\cite{Fermi-LAT:2013sme}.

IACTs  are sensitive to $\gamma$ rays in the Very-High Energy domain (VHE: $0.1 - 100$~TeV). They detect the faint Cherenkov light produced in the atmosphere by the secondary particles of the atmospheric air showers induced by $\gamma$ rays and cosmic rays. The properties of the image of the Cherenkov light on the telescope cameras are used to determine the characteristics of the incoming particle: its nature, its direction and its energy for instance. 
Current instruments, namely the High Energy Stereoscopic System (H.E.S.S.), the Major Atmospheric Gamma-ray Imaging Cherenkov telescope (MAGIC) or the Very Energetic Radiation Imaging Telescope Array (VERITAS) have searched for $\gamma$-ray signals from DM, and reported upper limits on the annihilation cross-section of DM particles~(see for example~\cite{Fermi-LAT:2013sme, HESS_DMsearch_WLM, HESS_DMsearch_DES, MAGIC_DMsearch_Triangulum, MAGIC_DMsearch_UrsaMajor, VERITAS_DMsearch_dsph}).
In particular, dwarf spheroidal galaxies (dSph) are ideal candidates to look at for indirect detection of DM because of their expected lack of non-thermal $\gamma$-ray emission processes~\cite{Winter2016}.
The next generation of IACT, the Cherenkov Telescope Array (CTA)~\cite{CTAConcept2013, CTAbook}, will consist of two arrays of hundreds of IACTs of different sizes, allowing a wide energy coverage from $\sim 20$~GeV to $\sim 300$~TeV over both hemispheres. With a sensitivity one order of magnitude better than current instruments, CTA will be in a unique position to discover a DM signal or, in the absence of it, to significantly improve the current DM limits.

A class of DM models that generically support TeV-scale or heavier DM are the portal-type models, where the dynamics of the dark sector is connected to the SM sector through a mediator. In this study, we investigate CTA sensitivity of a class of portal models in which the mediator is a scalar particle~\cite{Profumo:2010kp, Buckley:2014fba, Abdallah:2014hon, Feng:2014vea, Bishara:2015cha, Altmannshofer:2019wjb}. In such a model, DM can be either a bosonic or a fermionic particle. For concreteness, we focus on the inverse seesaw portal DM model (IS) introduced in~\cite{Pongkitivanichkul:2019cvm} and the inert doublet model (ID)~\cite{Deshpande:1977rw, LopezHonorez:2006gr, Goudelis:2013uca, Arcadi:2019lka}. The reason for this model specific approach instead of an effective approach is that the model parameters will allow us to project complementary limits from direct detection as well as collider limits on CTA sensitivity. We will first use collider limits and theoretical limits in order to choose suitable benchmarks. Then we use the CTA science analysis tools \texttt {ctools} to compute the expected CTA sensitivity to each model, adopting the Draco and Sculptor dSphs as representative targets for each hemisphere. Finally, we will compare the CTA sensitivity to the projected sensitivity of future direct detection experiments such as XENONnT~\cite{XENON:2020kmp} , by translating such limit to the usual indirect detection parameter space.

The outline of the paper is as follows. We introduce the two portal DM models in section~\ref{Portal_DM_models}, where also the benchmarks and specific DM spectra are discussed. The simulation of CTA data for the two targets, the Draco and Sculptor dSphs, is described in section~\ref{Methodology} as well as the method to derive the sensitivity curves for CTA. The results of the analysis are presented in section~\ref{Results}, where the CTA sensitivity is compared to direct detection limits from XENONnT. Our conclusions is drawn in section~\ref{Conclusion}.

\section{Portal dark matter models}
\label{Portal_DM_models}
In this section, we give a short review of the IS and ID scenarios. 

\subsection{Inverse seesaw model}
\label{ISmodel}
In the IS scenario, the DM particle is a Dirac fermion for which the dynamics of the DM sector is connected to the SM sector via the scalar mediator which is also responsible for neutrino mass generation. As a result, the neutrino mass scale sets the natural scale for the scalar mediator which, in turn, determines the scale of the DM particle. The smallness of neutrino mass dictates that DM must be of TeV scale. The potential for the scalar sector, $(\Phi,H)$, is given by
\begin{equation}
V(\Phi,H) = - \mu^{2}H^{\dagger}H + \lambda(H^{\dagger}H)^{2} - \frac{\mu_{\phi}^{2}}{2}\Phi^2 + \frac{\lambda_{\phi}}{4}\Phi^{4} + \frac{\lambda_{\phi H}}{2}\Phi^2 H^{\dagger}H, \label{eq:ISpot}
\end{equation}
where $\mu, \mu_{\phi}$ are the mass parameters and $\lambda, \lambda_{\phi}, \lambda_{\phi H}$ are the quartic couplings. The DM particle, $\chi$, is connected to the SM sector using the real scalar field, $\Phi$, as 
\begin{equation}
	\mathcal{L}_{\rm DM} = \Phi\overline{\chi}(G+i\tilde{G}\gamma^5)\chi + M\overline{\chi}\chi,
	\label{eq:darkmatter}
\end{equation}
where $M$ is the mass parameter for DM sector, $G (\tilde G)$ is a (pseudo-) scalar coupling. After symmetry breaking, we can expand the physical degrees of freedom as
\begin{eqnarray}
H = \frac{1}{\sqrt{2}}\begin{pmatrix} 0 \\ h' + v \end{pmatrix}, \quad\Phi = \phi' + v_{\phi},
\end{eqnarray}
where $v = 246$ GeV and $v_{\phi}$ are vacuum expectation values for the Higgs field and the real scalar field respectively. The mass of the DM particle can then be obtained by a chiral rotation and is equal to
\begin{equation}
    m_{\rm DM} = \sqrt{\frac{\left( \sqrt{2}M + v_{\phi} G\right)^2 + \left( v_{\phi} \widetilde{G}\right)^2}{2}}.
\end{equation}
The potential from Eq.~\ref{eq:ISpot} leads to the scalar mixing, i.e.,
\begin{align}
\begin{pmatrix} h \\ \phi \end{pmatrix} = \begin{pmatrix} \cos{\theta} & -\sin{\theta} \\ \sin{\theta} & \cos{\theta} \end{pmatrix} \begin{pmatrix} h' \\ \phi' \end{pmatrix}
\label{eq:scalarmixing}
\end{align}
where the mixing angle, $\theta$, together with $G$ and $\widetilde{G}$, control how strong the SM sector is connected to the DM sector. Since the mixing angle $\theta$ changes the coupling of the 125 GeV Higgs boson to other SM particles, the Higgs measurements from LHC~\cite{Khachatryan:2016vau,CMS:2018uag,ATLAS:2019nkf} can be translated to the constraint on the mixing angle, i.e., $|\cos\theta|\ge0.975$.

Due to the small mixing angle, the DM annihilation process is dominated by $\chi \overline{\chi}~\rightarrow~\phi \phi$, $h h \rightarrow j\overline{j} j' \overline{j'}$ where the final states are four SM particles, $j, j' = {\mu, \tau, c, b, t, W^{\pm}/Z}, g$. In the rest frame of the scalar, $\phi$ or h, the momenta of the SM particles are back to back. Hence the 4-momenta of the final states are isotropically distributed. Assuming isotropicity and boosting the momenta back to the DM center of mass frame, the energy of the final states $E_j$ ranges between $E_{\rm min}$ and $E_{\rm max}$ given by
\begin{equation}
E_{\rm max,min}^{\phi,h} = \frac{m_{\rm DM}}{2}\left( 1 \pm \sqrt{1-\left(\frac{m_{\phi,h}}{m_{\rm DM}}\right)^2}\sqrt{1-\left(\frac{2m_{j}}{m_{\phi,h}}\right)^2} \right).
\end{equation}
Averaging all possible directions, the differential probability of finding the SM particle with energy $E_{j}$ is
\begin{equation}
\left(\frac{dP}{dE_{j}}\right)_{\phi,h} = \frac{4}{\pi m_{\rm DM}}\frac{\sqrt{\left(1-\frac{m_{\phi,h}^2}{m_{\rm DM}^2}\right)\left(1-\frac{4 m_{j}^2}{m_{\phi,h}^2}\right) - \left( 1-\frac{2E_{j}}{m_{\rm DM}}\right)^2}}{\left(1-\frac{m_{\phi,h}^2}{m_{\rm DM}^2}\right)\left(1-\frac{4 m_{j}^2}{m_{\phi,h}^2}\right)}.
\end{equation} 
Finally, the $\gamma$-ray spectrum in the center of mass frame of the annihilating DM particle can be written as
\begin{equation}
    \left(\frac{dN_{j}}{dE_{\gamma}}\right)_{\phi,h\rightarrow j\overline{j}} = 2\int_{E_{\rm min}^{\phi,h}}^{E_{\rm max}^{\phi,h}} dE_{j} \left(\frac{dN_{j}}{dE_{\gamma}}\right)_{j}\left(\frac{dP}{dE_{j}}\right)_{\phi,h},
\end{equation}
where $\left(\frac{dN_{j}}{dE_{\gamma}}\right)_{j}$ is the $\gamma$-ray spectrum from the SM particle. To simplify the model further, we assume that $\phi$ has a negligible coupling to neutrinos such that it decays similarly to Higgs. Since the couplings to the SM particles of $\phi$ is modified by the same factor, we can assume that its branching ratios are similar to those of Higgs particle. 

The full spectrum coming from the DM annihilation is written as
\begin{equation}
    \frac{d\Phi}{dE_{\gamma}} = \sum_{j}\langle \sigma v\rangle_{\chi\overline{\chi} \rightarrow \phi\phi} {\rm Br}(\phi\rightarrow j\overline{j})\left(\frac{dN_{j}}{dE_{\gamma}}\right)_{\phi\rightarrow j\overline{j}} + \sum_{j}\langle \sigma v\rangle_{\chi\overline{\chi} \rightarrow hh} {\rm Br}(h\rightarrow j\overline{j})\left(\frac{dN_{j}}{dE_{\gamma}}\right)_{h\rightarrow j\overline{j}},
\end{equation}
where ${\rm Br}(X\rightarrow j\bar j)$ is a branching ratio of a mother particle $X$ into a pair of daughter particles $j\bar j$. The cross-section for DM particles annihilating into a pair of scalar bosons are given in Appendix~\ref{sec:AppendA}. Effectively, there are 6 parameters relevant to the DM direct and indirect detection, i.e, the DM particle mass, the scalar mediator mass, the vacuum expectation of the scalar field, the DM couplings and the mixing angle: 
\begin{equation}
\left(m_{\rm DM},m_{\phi},v_{\phi},G,\widetilde{G},\theta\right).
\end{equation}
The benchmark points are shown in Tab.~\ref{tab:benchmarkIS}.
\begin{table}[ht]
    \centering
    \begin{tabular}{|c|c|c|c|}
         \hline benchmark & $m_{\phi}$ & $v_{\phi}$ & $\sin^2(\theta)$  \\ \hline
         ISb1 & $m_{\phi} = m_{\rm DM} - 10$ GeV & $v_{\phi} = m_{\phi}$ & $2\times10^{-8}$ \\
         ISb2 & $m_{\phi} = m_{\rm DM} - 10$ GeV & $10^4$ GeV & $1\times10^{-8}$ \\
         ISb3 & $m_{\phi} = 0.9m_{\rm DM}$ & $v_{\phi} = m_{\phi}$ & $0.049375$ \\
         ISb4 & $m_{\phi} = 0.9m_{\rm DM}$ & $10^4$ GeV & $0.049375$ \\ \hline
    \end{tabular}
    \caption{Choice and relation between parameters for each benchmark. In all benchmarks, the DM particle mass, $m_{\rm DM}$, is treated as a free parameter. For ISb2 and ISb4, the DM particle mass is bounded below $\sim 10^5$ GeV due to the perturbativity of the scalar coupling. The values of the mixing angles in ISb3 and ISb4 are chosen from the saturation of the LHC constraints, i.e., $\cos\theta = 0.975$.}
    \label{tab:benchmarkIS}
\end{table}

\subsection{Inert doublet model}
\label{IDmodel}
For the ID model, an additional Higgs doublet, which does not participate in an electroweak symmetry breaking, is introduced. The physical neutral scalar or pseudoscalar component of this new Higgs doublet, whichever is the lightest, is a DM particle. As a result, the DM particle in this scenario is its own antiparticle. The DM dynamics in the ID scenario is connected to the dynamics of the SM Higgs boson. Compatibility with the observed DM relic density typically requires a heavy DM particle mass, $m_{\rm DM}\gtrsim600$ GeV~\cite{Goudelis:2013uca,Treesukrat:2019ahh}. The scalar sector of the ID model consists of the usual SM Higgs doublet, $\Phi$, and an additional doublet, $\Phi'$. The two doublets interact via a potential 
\begin{equation}
\begin{split}
	V(\Phi,\Phi') &= \mu^2\Phi^\dagger\Phi + \mu^{\prime2}\Phi^{\prime\dagger}\Phi' + \frac{\lambda_1}{2}(\Phi^\dagger \Phi)^2 + \frac{\lambda_2}{2}(\Phi^{\prime\dagger}\Phi')^2 \\
	&\quad + \lambda_3\Phi^\dagger\Phi\Phi^{\prime\dagger}\Phi' + \lambda_4 \Phi^\dagger\Phi' \Phi^{\prime\dagger}\Phi + \frac{\lambda_5}{2} \left((\Phi^{\prime\dagger} \Phi)^2+\text{h.c.}\right),
\end{split}
\label{eq:IDpot}
\end{equation}
where $\mu^{(\prime)}$ is the mass parameter and $\lambda_i$ is the quartic coupling.
The doublet $\Phi'$ does not participate in the Higgs mechanism, hence the name inert doublet. After electroweak symmetry breaking, the physical degrees of freedom in both doublets can be written as 
\begin{equation}
    \Phi = \frac{1}{\sqrt{2}}\begin{pmatrix}0\\v+h\end{pmatrix},\quad
    \Phi' = \frac{1}{\sqrt{2}}\begin{pmatrix}\sqrt{2}H^+\\H+iA\end{pmatrix},
\end{equation}
where $v$ = 246 GeV is the electroweak vacuum expectation value, $h$ is the SM Higgs boson observed at the LHC, $H^+$ is a charged scalar boson and $H$ and $A$ are neutral scalar bosons. The mass of the Higgs bosons are
\begin{equation}
    m_h^2=\lambda_1v^2,\quad 
	m_{H^+}^2 = \mu_2^2 + \frac{\lambda_3}{2}v^2,\quad 
	m_H^2 = m_{\phi^+}^2 + \frac{\lambda_4+\lambda_5}{2}v^2,\quad
	m_A^2 = m_{\rm DM}^2 - \lambda_5v^2.
\end{equation}
The lightest of the neutral particle $H$ and $A$ is the DM particle candidate. 

The phenomenology of the model depends on 5 unknown parameters: the 3 scalar masses $m_{H^+}$, $m_H$ and $m_A$ as well as two linear combinations of the quartic couplings typically taken to be $\lambda_2$ and $\lambda_{345} \equiv \lambda_3+\lambda_4+\lambda_5$. In the case where the DM particle is the $H$ particle, $\lambda_{345}$ is its coupling to the SM Higgs boson, $h$. If instead, $A$ is the DM particle, it couples to $h$ via $\bar\lambda_{345} \equiv \lambda_3+\lambda_4 - \lambda_5$. The DM-$h$ coupling, $\lambda_{345} (\bar\lambda_{345})$ plays a crucial role in determining the DM-nucleon scattering cross-section in direct detection experiments, DM annihilation cross-section in indirect detection experiments and the invisible decay of the SM Higgs boson in collider experiments.

Constraints on the ID model include theoretical bounds from stability of the electroweak vacuum~\cite{Ivanov:2006yq} and unitarity of the 2-2 scattering amplitudes~\cite{Ginzburg:2003fe}. These translate to constraints on the quartic couplings $\lambda$'s as follow:
\begin{equation}
\begin{split}
    \lambda_1>0,\quad\lambda_2>0,\quad\lambda_3>-\sqrt{\lambda_1\lambda_2},\quad
	\lambda_3+\lambda_4-|\lambda_5|>-\sqrt{\lambda_1\lambda_2},\\
	\left|\lambda_3+2\lambda_4\pm3\lambda_5\right|\le8\pi,\quad3(\lambda_1+\lambda_2)+\sqrt{9(\lambda_1-\lambda_2)^2+4(2\lambda_3+\lambda_4)^2}\le16\pi.
\end{split}
\end{equation}
These constraints put an upper limit on how heavy a DM particle can be. 

From the experimental side, DM particles can be searched for at a particle collider such as the LHC. However, for heavy DM scenario ($m_{\rm DM}\gtrsim 1$ TeV), collider search loses sensitivity because these DM particles cannot be produced efficiently~\cite{Arcadi:2019lka}. Indeed, the sensitivity of direct detection experiments depends on the flux of DM particles passing through the detector. Hence, since the flux decreases with increasing DM particle mass, direct detection experiments lose sensitivity for heavy DM particles. For indirect detection experiments, the sensitivity for heavy DM particles is limited by the telescope ability to detect high energy $\gamma$ rays.

\begin{table}[ht]
    \centering
    \begin{tabular}{|c|c|c|}
         \hline benchmark & $\Delta^0$ [GeV]& $\Delta^+$ [GeV]  \\ \hline
         IDb1 & 2 & 7 \\
         IDb2 & 0.2 & 0.7  \\ \hline
    \end{tabular}
    \caption{Benchmark scenarios for the ID model. For a given value of $\Delta^0$, $\Delta^+$ and $m_{\text{DM}}$, $\lambda_{345}$ is chosen so that the correct DM relic density is reproduced.}
    \label{tab:benchmarkID}
\end{table}

The ID model contains 7 real parameters as can be seen in Eq.~\ref{eq:IDpot}. Two of the parameters are fixed by the 125 GeV Higgs boson mass and the electroweak vacuum expectation value. The remaining 5 free parameters are taken to be the DM particle mass $m_{\rm DM}$, the mass splitting between the DM particle and its neutral cousin $\Delta^0$, the mass splitting between the DM particle and its charged cousin $\Delta^+$, $\lambda_{345}$ and $\lambda_{2}$. In this study, we consider two benchmark scenarios for $\Delta^0$ and $\Delta^+$ shown in Tab.~\ref{tab:benchmarkID}. In each benchmark scenario, for a given value of $m_{\rm DM}$, $\lambda_{345}$ is chosen so that the correct relic abundance of DM particles is produced. Note that the parameter $\lambda_2$ plays no role in DM phenomenology. 


\section{Methodology}
\label{Methodology}
\subsection{Targets and CTA simulation}
\begin{figure}
    \centering
    \includegraphics[scale=0.5]{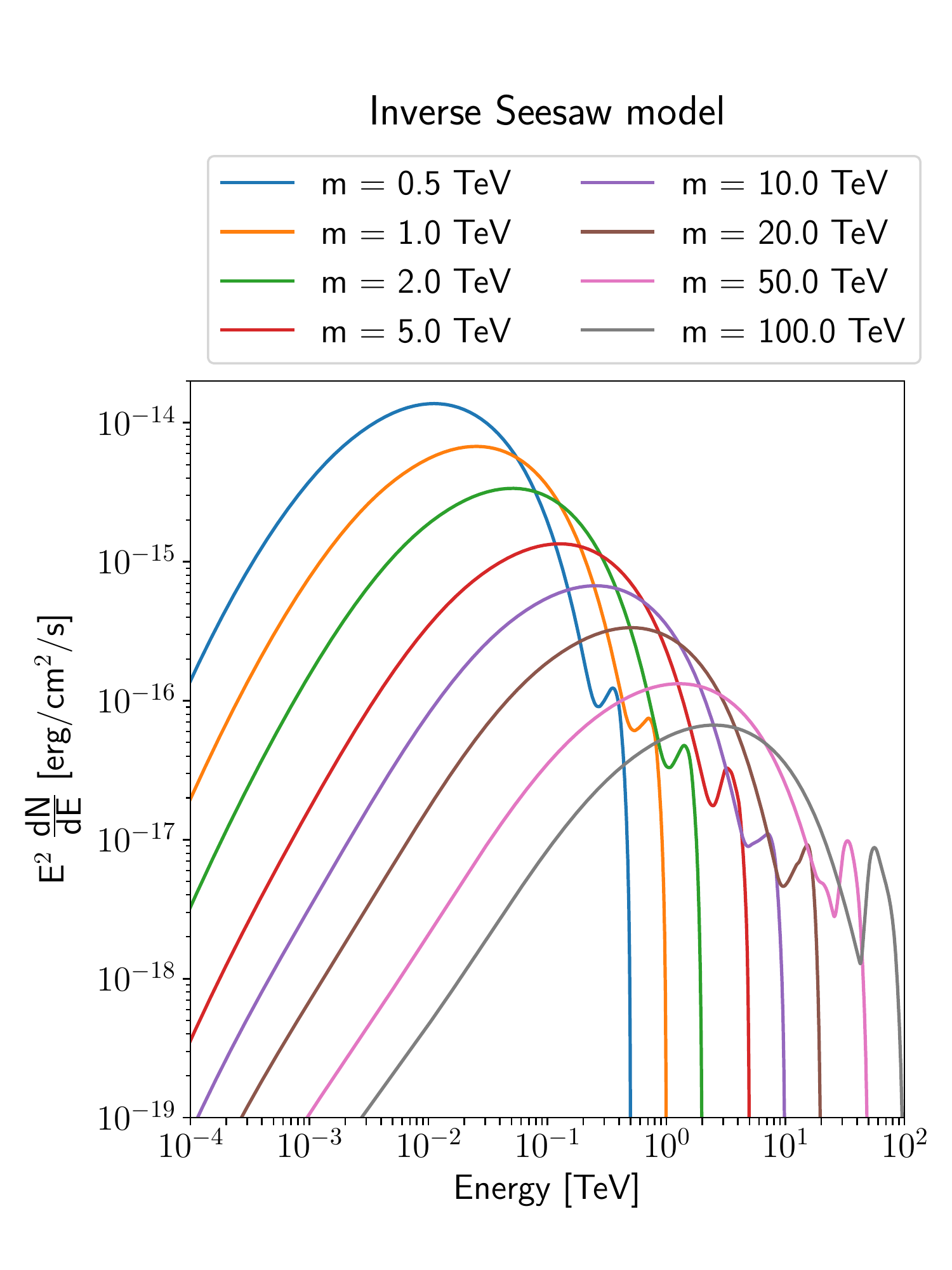}
    \includegraphics[scale=0.5]{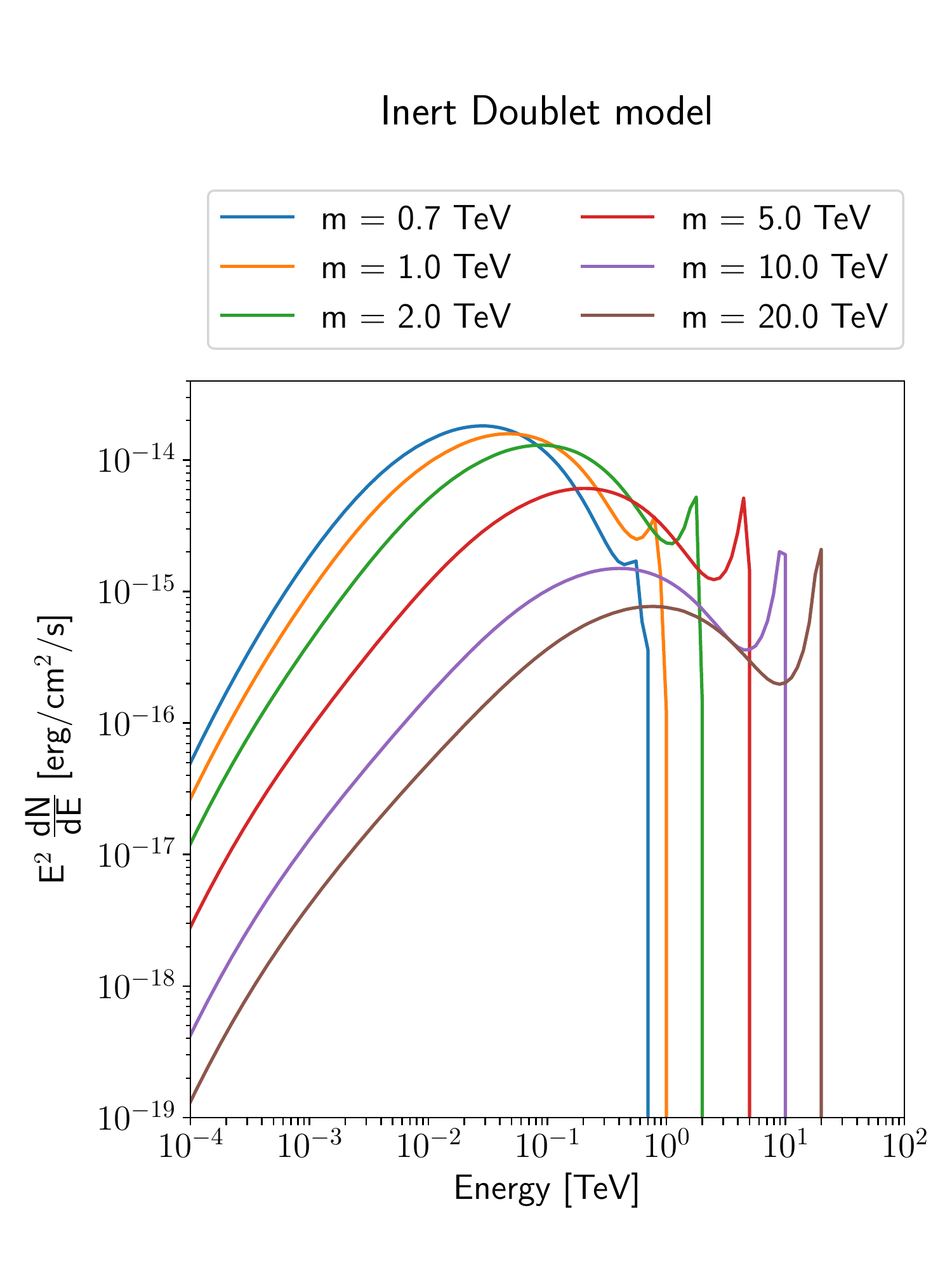}
    \caption{Spectral energy distribution for the inverse seesaw model with benchmark scenario ISb1 (left) and the inert doublet model with benchmark IDb1 (right), adopting the Draco $J$ factor  $J~=~1.42 \times 10^{19}$ GeV$^2$ cm$^{-5}$, for different DM  particle masses. }
    \label{fig:IDspectrum_draco}
\end{figure}

The source differential DM flux is given by
\begin{equation}
    \frac{d\phi}{dE}=\frac{1}{8\pi}\frac{J}{m_{\rm DM}^{2}} \sum_{i}\frac{dN_i}{dE}\left<\sigma v\right>_i
\end{equation}
where $J=\int \rho^2 dr$ is the astrophysical factor corresponding to the DM density squared integrated along the line of sight. The overall factor $1/8\pi$ comes from the combination of solid angle, the assumption that there are as many DM particles as their antiparticles and the fact that the spectrum from a SM particle and its antiparticle are identical. $\frac{dN_i}{dE}$ is the differential photon number density from DM annihilation in the channel $i$. This equation applies in the case where the DM particle is not its own antiparticle as it is the case for the IS model. If the DM particle is its own antiparticle, as in the case of the ID model, the flux should be multiplied by a factor 2. Examples of $\gamma$-ray spectra for both the IS and the ID models from DM annihilation inside Draco are shown in Fig.~\ref{fig:IDspectrum_draco} for various DM particle masses. Note that for the ID model, the DM particle mass must be between 0.7 and 23 TeV. Hence, the spectrum is shown up to 20 TeV.

Two dwarf spheroidal galaxies are used as targets: Draco in the northern hemisphere, and Sculptor in the southern hemisphere. 
Both are classical DM targets, extensively studied before, and the uncertainties on the DM flux are expected to be much lower compared to other DM targets. For these resaons, they are 
suited to test the IS and the ID model presented in sections~\ref{ISmodel} and \ref{IDmodel}. 
As a typical exposure time~\cite{CTAbook}, 300 hours of observations of these two targets are simulated using the \texttt{ctools} software package~\cite{ctools} and adopting CTA-North array for Draco and the CTA-South array for Sculptor.  Tab.~\ref{target_parameters} gathers the useful parameters for these two targets. 
Gamma-ray events with energy between 30~GeV and 100~TeV are generated from each target source with \texttt{ctobssim} assuming a point-like source, using the corresponding instrument response functions (IRF~\cite{irf-prod3}), and a pointing offset of 0.5\textdegree\ from the source coordinates.

\begin{table}[ht!]
 \centering
\begin{tabular}{|c|c|c|c|c|}
\hline
 \multirow{2}*{Target} &  \multirow{2}*{(RA,Dec) [\textdegree]} & \multirow{2}*{IRF} &  \multirow{2}*{\textit{J} $[10^{18} \frac{\text{GeV}^2}{\text{cm}^5}]$} \\
                       &                                        &                    &                  \\
\hline
Draco & (260.05, 57.915) & North\_z20\_50h & 14.2 \\
Sculptor & (15.0375, -33.7092) & South\_z20\_50h & 3.56 \\
\hline
\end{tabular}
\caption{Parameters of the two dwarf spheroidal galaxies used as targets in this study. The IRFs correspond to the \texttt{prod3b\_v2} publicly released by CTA~\cite{irf-prod3}. The $J$ factors are taken from~\cite{2015DMinDsph}.}
\label{target_parameters}
\end{table}

\subsection{Sensitivity curves and upper limits}

To derive the CTA sensitivity for both models and for both targets, we closely follow the methodology in~\cite{Aguirre_Santaella_2020}. For each tested DM particle mass, from 30 GeV to 100 TeV, we use \texttt{ctlike} to calculate the Test Statistic (TS), which compares the likelihood that a signal $s$ is present, $\mathcal{L}(M_{s+b})$, against the hypothesis that there is background $b$ only, $\mathcal{L}(M_{b})$:
    \begin{equation}
        TS=2(\ln\mathcal{L}(M_{s+b})-\ln\mathcal{L}(M_b)).
    \end{equation}
The simulated $\gamma$-ray flux predicted by each model is boosted by a factor $f$, which is equivalent to increasing the cross-section, until the signal is detected with a $TS$ of $25\pm0.25$.

In the case where no signal is detected with $\Delta TS \ge 25$ within 300 hours of simulated observations, we also calculate the upper limit on the flux, $f_{\text{ulm}}$, using \texttt{ctulimit}. We consider the 95\% confidence level (CL) upper limit and derive the upper limit boost factor, $f_{\text{ulm}}$, by
    \begin{equation}
        f_{\text{ulm}} =\frac{\phi_{\text{ulm}} }{\int_{E_{\rm min}}^{m_{\rm DM}} \frac{d\phi}{dE}dE},
    \end{equation}
where $E_{\rm min} = 30$ GeV and $m_{\rm DM}$ the mass of the DM particle. 
Finally, the $f_{\text{ulm}}$ is translated to the 95\% CL upper limit on the DM annihilation cross-section. In both cases we performed 50 realisations and calculated the mean and standard deviation. 


\section{Results}
\label{Results}
\subsection{Detection prospects and upper limit to the annihilation cross section}

\begin{figure}[!ht]
    \centering
    \begin{tabular}{cc}
        \includegraphics[width=.45\textwidth]{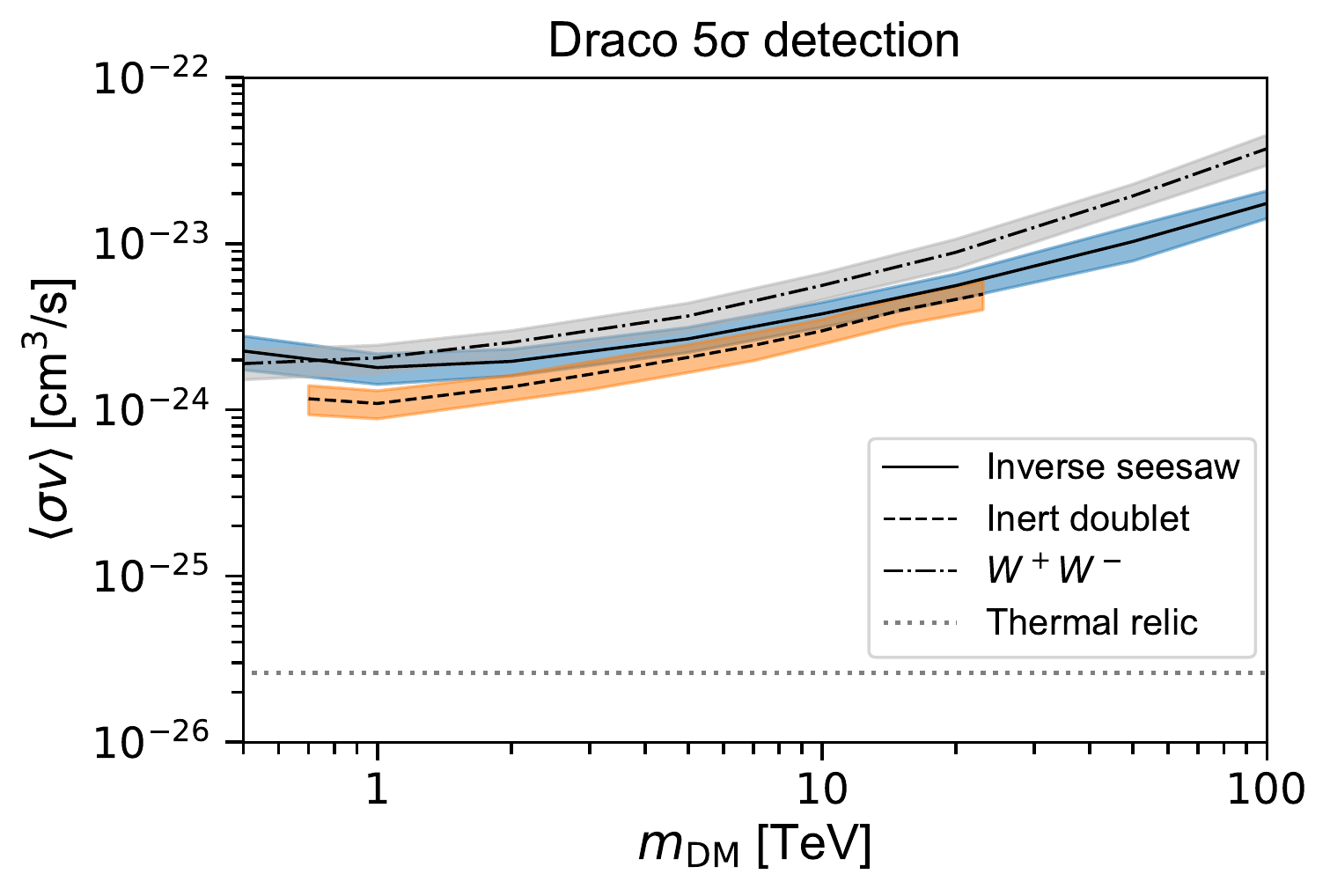} & \includegraphics[width=.45\textwidth]{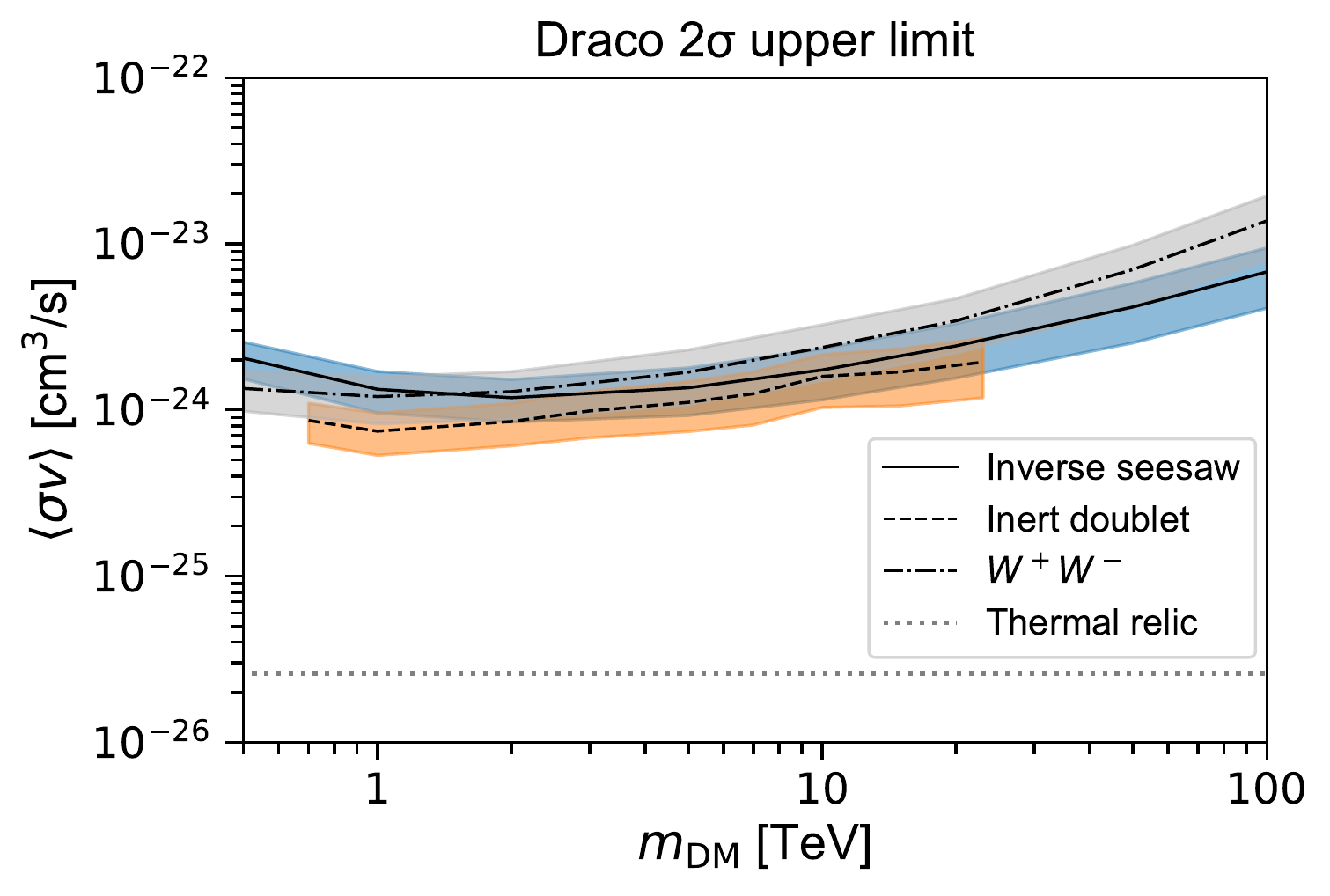}  \\
        \includegraphics[width=.45\textwidth]{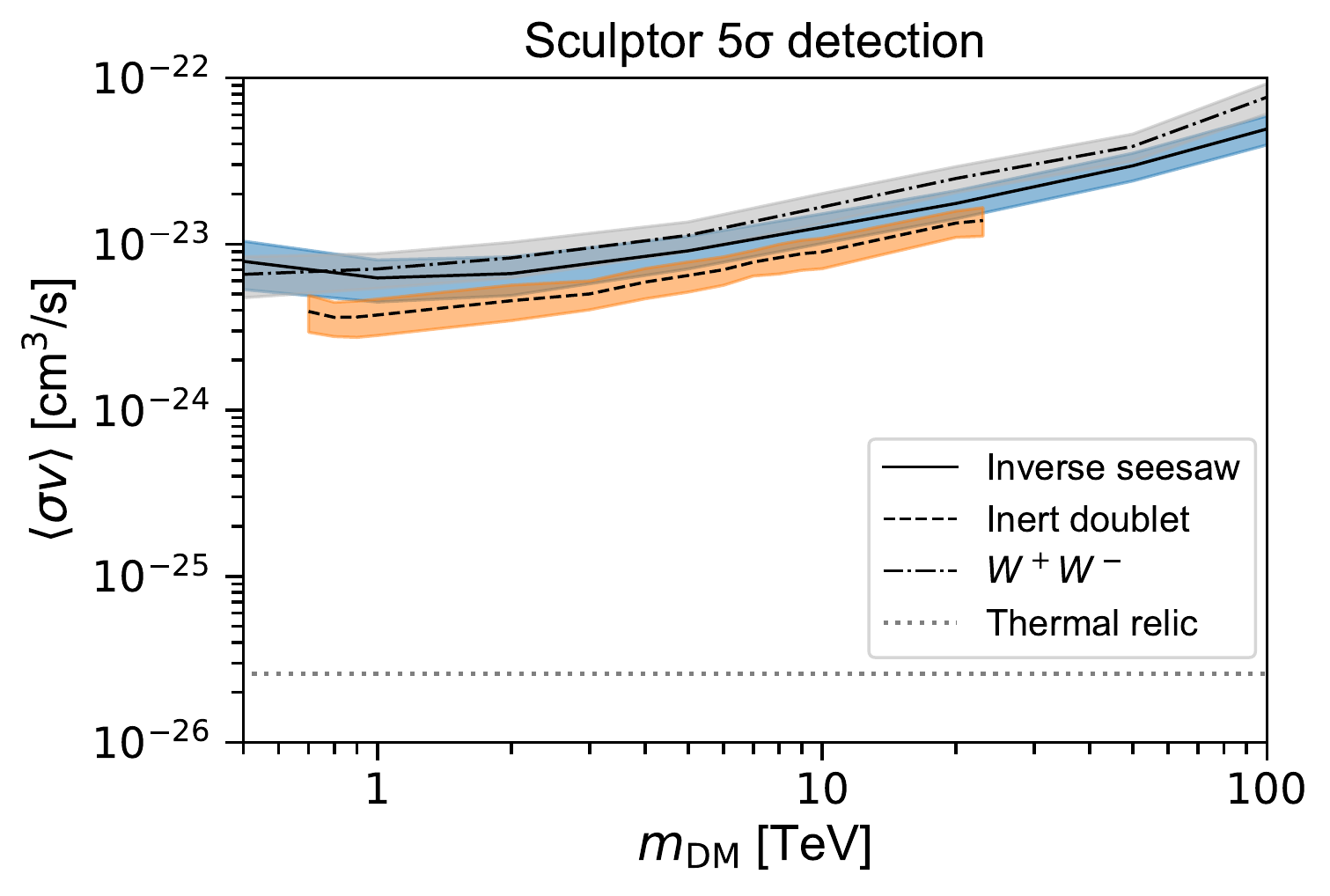} & \includegraphics[width=.45\textwidth]{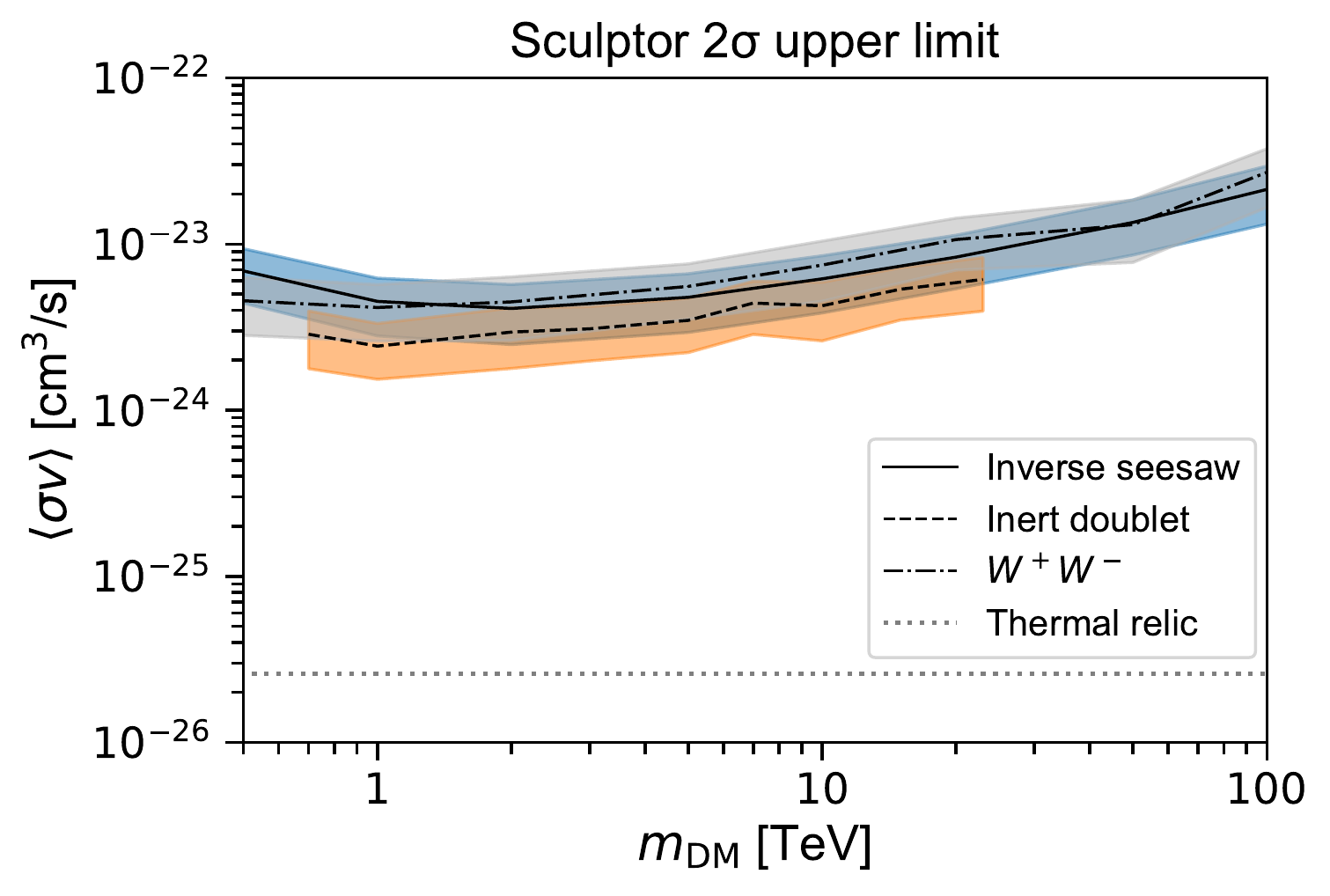} 
    \end{tabular}
    \caption{Annihilation cross-section as a function of the DM particle mass for the IS model (blue curves) and the ID model (orange) in the case of Draco at the top and Sculptor at the bottom. The cross-section for the SM annihilation channel $W^{+}W^{-}$ is also shown for comparison (green curve). The shaded bands represent the 1$\sigma$ uncertainties.  
    }
    
    \label{cta-results}
\end{figure}

Using 300 hours of simulated CTA observations of the Draco and the Sculptor dSph galaxies, the value of the cross-section that would be needed for the 5$\sigma$ discovery limit of the DM annihilation signal from Draco (Sculptor) is shown in the top left (right) panel of Fig.~\ref{cta-results} for both the IS and the ID models.  Similarly, the 95\% CL exclusion limits for both targets, in case no signal was found, are shown in the bottom panels of Fig.~\ref{cta-results}. For comparison, we also provide our estimate for a simplified scenario where DM only annihilates to $W^{+}W^{-}$. 

For both targets, the IS model provides both an easier discovery and better exclusion limits compared to the $W^{+}W^{-}$ scenario for $m_{\rm DM}\gtrsim $ 1 TeV.  In the IS model, DM particles annihilate into a pair of secondary particles ($\phi$ and $h$), which subsequently decay into a pair of SM particles ($\phi,h\to W^{+}W^{-}$, $t\bar t$). This two-step process accounts for most of the $\gamma$-ray spectrum produced. As a result, for DM particle masses below 1 TeV, corresponding to 0.5 TeV energy of the SM particle producing $\gamma$ rays, the majority of the $\gamma$-ray spectrum is below the CTA sensitivity threshold. On the other hand, the $t\bar t$ channel, which gives a stronger $\gamma$-ray spectrum at higher energy, provides the strongest limit for the IS model compared to the $W^{+}W^{-}$ spectrum in the region of $m_{\rm DM}\gtrsim$~1~TeV. 

The ID model provides both the best prospects for discovery and most stringent exclusion bounds in case no signal is detected. This is due to the fact that for $m_{\rm DM}\gtrsim600$ GeV, consistency with the DM relic abundance is achieved via the co-annihilation mechanism~\cite{Griest:1990kh}. It allows the DM annihilation cross-section to be larger than the nominal thermal relic cross-section $\langle\sigma v\rangle\simeq2.6\times10^{-26}$~cm$^3$/s. For the benchmark scenario considered, internal consistency dictates that the DM particle mass must lie in a region between 0.7 and 23 TeV. As a result, the energy range that can be probed with the ID model is smaller than with the IS model.

\subsection{Constraint from direct detection experiments}
In this subsection we compare the projected CTA sensitivity to a DM signal from dSph galaxies to that of direct detection experiments. For definiteness, we take the XENONnT projection~\cite{XENON:2020kmp} as a representative of direct detection limits. Other planned direct detection experiments such as DarkSide-20k~\cite{DarkSide-20k:2017zyg}, LUX-ZEPLIN~\cite{LUX-ZEPLIN:2018poe} and PandaX-4T~\cite{PandaX:2018wtu} are expected to provide comparable sensitivity.

\begin{figure}[ht]
    \centering
        \includegraphics[width=0.55\textwidth]{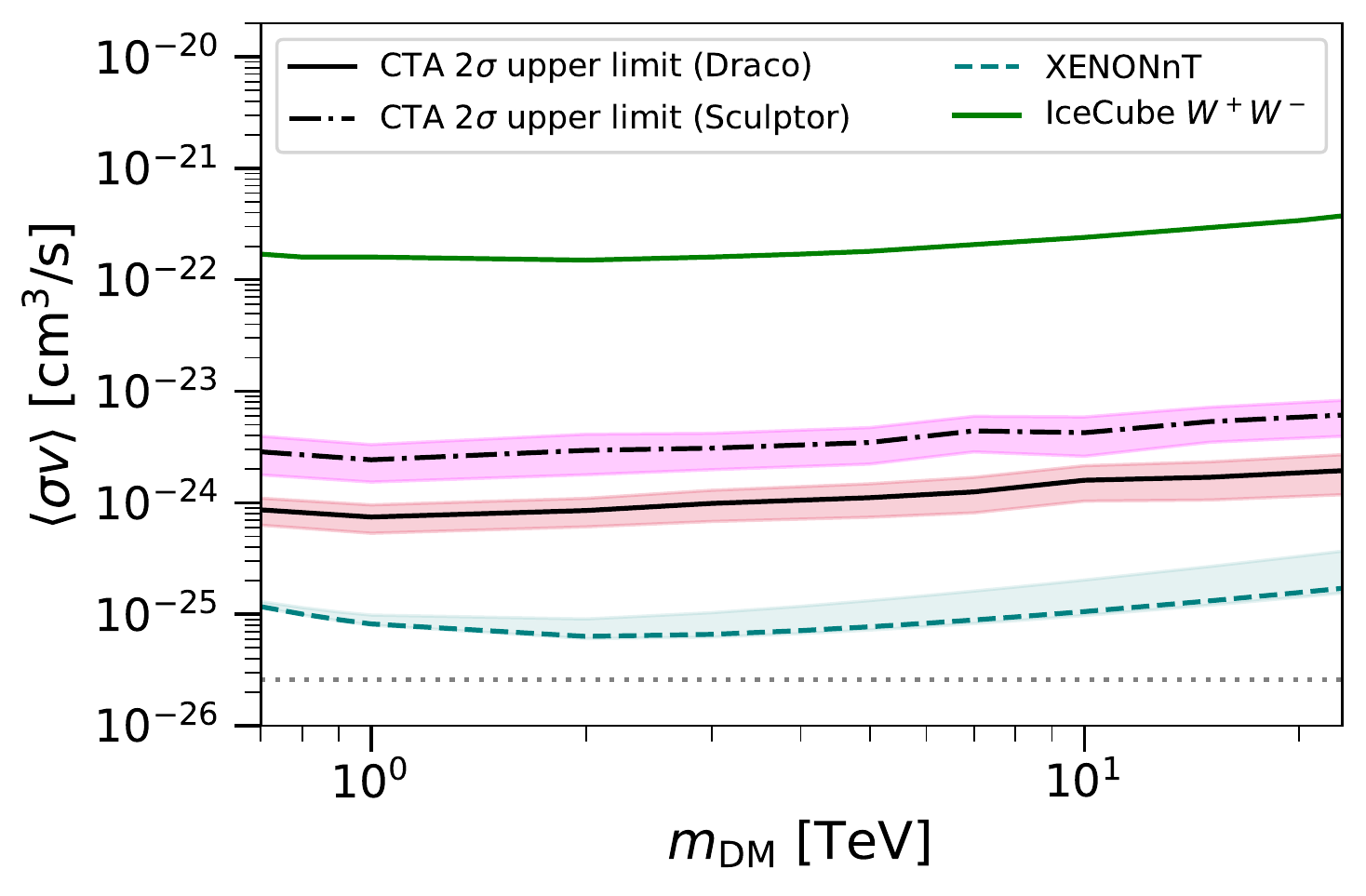}  
    \caption{Comparison between annihilation cross sections in the Inert Doublet model (solid for Draco; dot-dashed for Sculptor), the IceCube upper limit (solid green) and the one translated from XENONnT (light blue); see text for details on the latter. Shaded bands are $1\sigma$ uncertainties.}
    \label{fig:combined}
\end{figure}

We translate the XENONnT constraint on the spin-independent DM-nucleon scattering cross-section into a constraint on DM annihilation cross-section. The direct detection cross-sections for IS and ID models can be found in Appendix~\ref{sec:AppendB}. Since XENONnT only provides an upper bound up to $m_{\rm DM}=1$ TeV, we perform a linear extrapolation to get an upper bound for heavier DM masses. Fig.~\ref{fig:combined} shows a comparison between the projected CTA and XENONnT sensitivities. For the benchmark scenarios considered here, XENONnT provides better constraints for the ID model than CTA. 
However the XENONnT limit is not strong enough to exclude the parameter space relevant for the IS model. This can be seen from Eq.~\ref{eq:ddxsec}. Due to the strong suppression of the mixing angle, projection limits on spin-independent cross-section from XENONnT, $\sigma_{SI} \sim 10^{-47}$ ${\rm cm}^{2}$, implies that the excluded values of the DM coupling to the mediator are generically larger than the perturbativity limit, i.e., $G^2 + \tilde{G}^2 \gtrsim 10^2$ for $m_{\rm DM} \gtrsim 10^3$ GeV.

For the ID model, we also compare the CTA sensitivity to the exclusion limit ($90\%$ CL) of one year data from IceCube \cite{IceCube:2014rqf}. 
We find that the projected CTA sensitivities for both Draco and Sculptor are stronger than IceCube upper limit by roughly 2 orders of magnitude. On the other hand, the typical neutrino production cross-section from the IS model is expected to be of the order $~10^{-31}$ cm$^3$/s \cite{Pongkitivanichkul:2019cvm}, which is much lower than the current constraints from IceCube.

\begin{figure}
    \centering
    \includegraphics[width=0.95\textwidth]{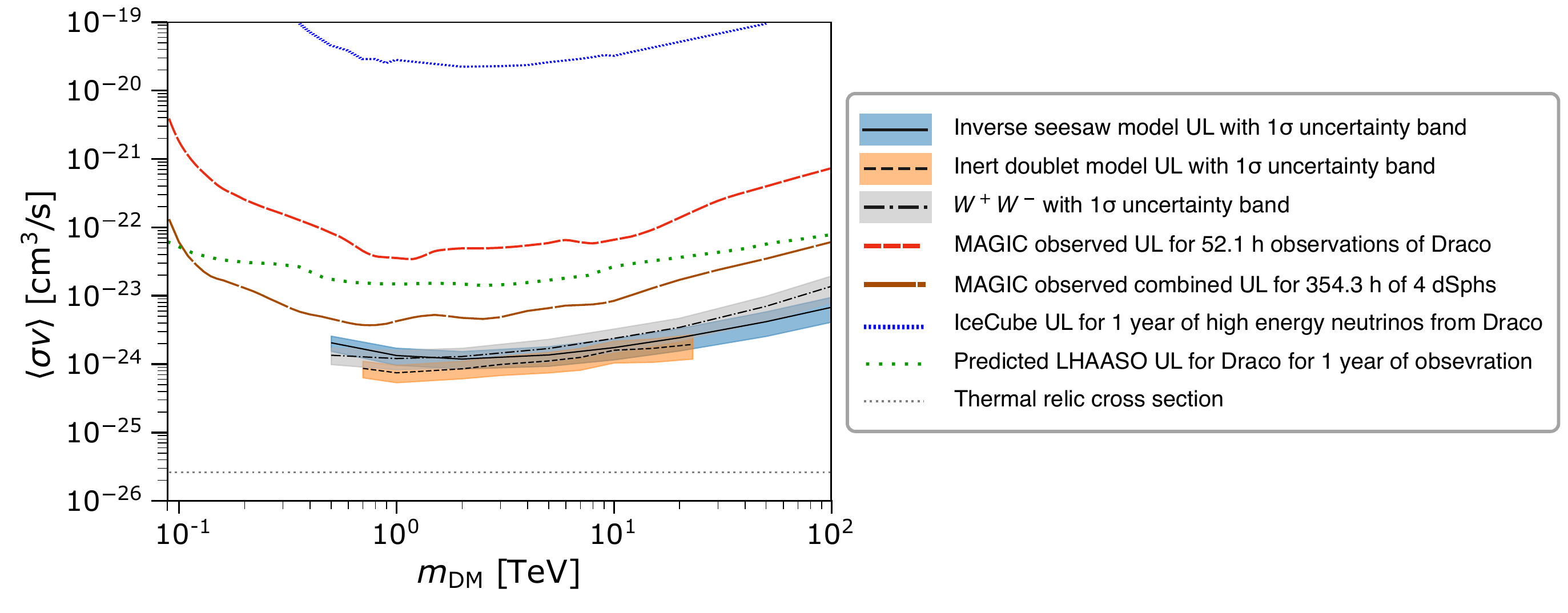}
    \caption{ Comparison of the 95\% CL ULs on the DM annihilation cross-section for the IS and ID model and for the $W^{+}W^{-}$ channel, with 95\% CL ULs derived by the MAGIC collaboration for the $W^{+}W^{-}$ channel~\cite{Dsph_DM_search_MAGIC}. The blue dashed line represents the 90\% CL UL from IceCube for the $W^{+}W^{-}$ channel in the direction of Draco~\cite{IceCube_draco_UL} and the green dashed line shows the predicted upper limit at 95\% confidence level after 1 year of observation with LHAASO for the $W^{+}W^{-}$ channel~\cite{LHAASO_dsph_UL}. 
    }
    \label{fig:MAGIC}
\end{figure}

\section{Discussion and Conclusion}
\label{Conclusion}
In this paper, we studied the detectability of DM particles in the context of portal-type DM models, namely the inverse seesaw model and the inert doublet model. CTA is a particularly suited instrument thanks to its sensitivity in the TeV energy range, where we expect DM signal from the two models considered in this study. First, we derived the expected cross-section for a 5$\sigma$ detection using the Draco and the Sculptor dwarf spheroidal galaxies as astrophysical targets. Unfortunately, the DM annihilation cross-section required for a discovery is 2 to 3 orders of magnitude higher than the thermal relic cross-section. Therefore, it seems unlikely that CTA will detect DM particles from the IS and ID models. Hence, for the case of no signal, we derived the 95\% CL upper limits on the annihilation cross-section. These limits turn out to be weak as they are also 2-3 orders of magnitude above the typical thermal relic cross-section. 
In Fig.~\ref{fig:MAGIC}, we compare them with the upper limits for the $W^{+}W^{-}$ channel published by the MAGIC collaboration for $\sim52$~hours of Draco observations and for the combined analysis of four dSph galaxies (Segue I, Ursa Major II, Draco and Coma Berenices) leading to $\sim350$~hours of combined observations~\cite{Dsph_DM_search_MAGIC}. As expected, the upper limit of CTA is more than one order of magnitude stronger than MAGIC in the case of Draco alone, and $\sim4$ times stronger in the case of the combined limit. The upper limit at 90\% confidence level from IceCube~\cite{IceCube_draco_UL} using high energy neutrinos over 340 days coming from the direction of Draco for the $W^{+}W^{-}$ channel is also shown, as well as the upper limit at 95\% confidence level predicted after 1 year of observation with LHAASO for the $W^{+}W^{-}$ channel~\cite{LHAASO_dsph_UL}. 
For the ID model, we also compare in Fig.~\ref{fig:combined} our simulated CTA exclusion limit against the projected XENONnT limit, which is roughly 1 order of magnitude stronger. A similar comparison for the IS model is not possible since the DM-nucleon cross-section is highly suppressed due to a small mixing between the mediator $\phi$ and the Higgs boson $h$, which is tightly constrained by the 125 GeV Higgs boson measurements at the LHC.  

We note that the CTA sensitivity could be improved by either astrophysical boosting effects, such as black-hole induced clamping \cite{Gammaldi:2016uhg,Gondolo:1999ef} or subhalo boosting \cite{Sanchez-Conde:2013yxa,Moline:2016pbm,Ando:2019xlm}, or by Sommerfeld enhancement of the annihilation cross-sections~\cite{Hisano:2004ds,Lattanzi:2008qa}. 
Another improvement could come from a combined analysis of multiple dSph observations. Indeed, the latter can increase the statistics and improve the upper limit by at least a factor 2. For example, the combination of the observations from 20 dSph galaxies, recently performed by the Fermi-LAT, HAWC, H.E.S.S., MAGIC, and VERITAS collaborations using a joint maximum likelihood approach, leads to upper limits that are 2 to 3 times stronger than using individual instruments~\citep{Combined_DM_searches}.

It must be also noted that direct detection constraints are sensitive to the local DM density, $\rho_0$. Recent studies of the DM distribution in the Milky Way suggest that 0.4 GeV/cm$^3\lesssim\rho_0\lesssim0.7$ GeV/cm$^3$ at 95\% CL~\cite{Benito:2019ngh,Benito:2020lgu} with the best fitted value $\rho_0\simeq0.57$ GeV/cm$^3$. This would result in a more constraining direct detection upper limit. However, in our analysis, we follow XENONnT by adopting the standard value $\rho_0=0.3$ GeV/cm$^3$. As a result, our projected constraints from direct detection are conservative.

Although the CTA sensitivity for portal-type DM models is weak for Draco and Sculptor, the Galactic centre (GC) can be the next promising target to test the model. Recent results for the sensitivity of CTA to a DM signal from the GC, with both cusp and core profiles, have shown that CTA will be able to reach the thermal cross-section for TeV-scale DM \cite{CTACollaboration2021}.
In the near future, we will test IS and ID models in the GC following the method in \cite{CTACollaboration2021}, which also considered the interstellar emission model, known and unknown sources in the GC area and the Fermi bubbles.\\

\section*{Acknowledgments}
This work was conducted in the context of the Dark Matter and Exotic Physics Working Group of the CTA Consortium. 
This research made use of \texttt{ctools}~\cite{ctools}, a community-developed analysis package for Imaging Atmospheric Cherenkov Telescope data, based on \texttt{GammaLib}, a community-developed toolbox for the scientific analysis of astronomical $\gamma$-ray data.
The CTA instrument response functions provided by the CTA Observatory (version \texttt{prod3b-v2} \cite{irf-prod3}) were also used.
The sensitivity results were made possible by the Chalawan cluster, a High-Performance Computing (HPC) Cluster at National Astronomical Research Institute of Thailand (NARIT).
CP is supported by Research Grant for New Scholar, Office of the Permanent Secretary, Ministry of Higher Education, Science, Research and Innovation under contract no. RGNS 64-043. CP has also received funding support from the National Science, Research and Innovation Fund (NSRF). AJB acknowledges the support from Chulalongkorn University’s CUniverse (CUAASC) grant and from the Program Management Unit for Human Resources and Institutional Development, Research and Innovation, NXPO (grant number B16F630069).
The work of PU was supported in part by the Mid-Career Research Grant from National Research Council of Thailand under contract no.~N42A650378 and the Thailand Toray Science Foundation. MW acknowledges the support of Kasetsart University Research and Development Institute, KURDI. 
The work of AAS and MASC was supported by the grants PGC2018-095161-B-I00 and CEX2020-001007-S, both funded by MCIN/AEI/10.13039/501100011033 and by ``ERDF A way of making Europe''. MASC was also supported by the Atracci\'on de Talento contract no. 2020-5A/TIC-19725 granted by the Comunidad de Madrid in Spain. The work of AAS was also supported by the Spanish
Ministry of Science and Innovation through the grant FPI-UAM 2018.

\appendix
\section{Analytic expressions for cross-sections}

In this appendix we provide analytic expressions for DM self-annihilation cross-sections, relevant for indirect detection experiments, and the DM-nucleon cross-sections, responsible for direct detection experiments.
\subsection{Self-annihilation cross-sections}
\label{sec:AppendA}
For the IS model, annihilation into the $\phi\phi$ and $hh$ final states,  responsible for most of the annihilation cross-sections, are given by
\begin{align}
	\langle\sigma v\rangle_{hh} 
	&= \frac{\beta_\chi^2G^2+\tilde{G}^2}{32\pi}\sqrt{1-\frac{m_h^2}{m_{\rm DM}^2}} \left(\frac{4m_{\rm DM} G s_\theta^2}{2m_{\rm DM}^2-m_h^2} -\frac{s_\theta\lambda_{hhh}}{4m_{\rm DM}^2-m_h^2} - \frac{c_\theta\lambda_{\phi hh}}{4m_{\rm DM}^2-m_\phi^2}\right)^2, \\
	\langle\sigma v\rangle_{\phi\phi} 
	&= \frac{\beta_\chi^2G^2+\tilde{G}^2}{32\pi}\sqrt{1-\frac{m_\phi^2}{m_{\rm DM}^2}} \left(\frac{4m_{\rm DM} G s_\theta^2}{2m_{\rm DM}^2-m_\phi^2} -\frac{s_\theta\lambda_{\phi \phi h}}{4m_{\rm DM}^2-m_h^2} - \frac{c_\theta\lambda_{\phi\phi\phi}}{4m_{\rm DM}^2-m_\phi^2}\right)^2.
\end{align}
where $\beta_\chi \equiv \sqrt{1-\frac{4m_{\rm DM}^2}{s}}$ and the couplings are given by
\begin{align}
	\lambda_{h h h} &= -\frac{6 s^3_{\theta} \left(m_{\phi}^2c^2_{\theta} + m_h^2s^2_{\theta} \right)}{v_{\phi}}-\frac{6 c^3_{\theta} \left(m_{h}^2c^2_{\theta} + m_{\phi}^2s^2_{\theta} \right)}{v}\nonumber \\ 
	& \quad-\frac{\left(m_{\phi}^2 - m_h^2\right)c_{\theta}s_{\theta}}{vv_{\phi}}\left(3v_{\phi}c^2_{\theta}s_{\theta} + 3vc_{\theta}s^2_{\theta}\right),\\
	\lambda_{\phi h h} &=-3s_{2\theta}\left[m_{\phi}^2\left(\frac{c_\theta s^2_{\theta}}{v_\phi} + \frac{c^2_{\theta}s_\theta}{v} \right) + m_{h}^2\left(\frac{c^3_{\theta}}{v_\phi} + \frac{s^3_\theta}{v} \right)\right]\nonumber\\
	& \quad -\frac{\left(m_{\phi}^2 - m_h^2\right)c_{\theta}s_{\theta}}{vv_{\phi}} \left(v_{\phi}c^3_{\theta} +2vc^2_{\theta}s_{\theta} - 2v_{\phi}c_{\theta}s^2_{\theta} - vs^3_{\theta}\right), \\
	\lambda_{h \phi \phi} &= -3s_{2\theta}\left[m_{\phi}^2\left(\frac{c^3_{\theta}}{v_\phi} + \frac{s^3_\theta}{v} \right) + m_{h}^2\left(\frac{c_\theta s^2_{\theta}}{v_\phi} + \frac{c^2_{\theta}s_\theta}{v} \right)\right]\nonumber\\
	& \quad -\frac{\left(m_{\phi}^2 - m_h^2\right)c_{\theta}s_{\theta}}{vv_{\phi}}\left(v_{\phi}s^3_{\theta} -2vc_{\theta}s^2_{\theta} - 2v_{\phi}c^2_{\theta}s_{\theta} + vc^3_{\theta}\right), \\
	\lambda_{\phi \phi \phi} &= -\frac{6 c^3_{\theta} \left(m_{\phi}^2c^2_{\theta} + m_h^2s^2_{\theta} \right)}{v_{\phi}}+\frac{6 s^3_{\theta} \left(m_{h}^2c^2_{\theta} + m_{\phi}^2s^2_{\theta} \right)}{v}\nonumber\\
	& \quad -\frac{\left(m_{\phi}^2 - m_h^2\right)c_{\theta}s_{\theta}}{vv_{\phi}}\left(3v_{\phi}c_{\theta}s^2_{\theta} - 3vc^2_{\theta}s_{\theta}\right).
\end{align}

For the ID model, the relevant annihilation cross-sections in the case where $H$ is a DM particle, are given by
\begin{align}
    \langle\sigma_{f\bar f} v\rangle &= \frac{N_c}{4\pi}\frac{\lambda^2_{345}m_f^2}{(4m_\chi^2-m_h^2)^2}(1-r_f)^{3/2},\\
    \langle\sigma_{hh} v\rangle &= \frac{1}{64\pi}\frac{\lambda^2_{345}}{m_\chi^2}\left[1 + \frac{3m_h^2}{4m_\chi^2-m_h^2} + \frac{2\lambda v^2}{m_h^2-2m_\chi^2}\right]^2\sqrt{1-r_h}\,,\\
    \langle\sigma_{VV} v\rangle &= \frac{m_{DM}^2}{2\delta_V\pi v^4}\left[\left(1+\frac{\lambda v^2}{4m_\chi^2-m_h^2}\right)^2\left(1-r_V+\frac34r_V^2\right) +\frac{4m_\chi^4}{(m_\phi^2+m_\chi^2-m_V^2)^2}(1-r_V)^2\right.\nonumber\\
    &\qquad\left.-\frac{2m_\chi^2}{(m_\phi^2+m_\chi^2-m_V^2)}\left(1+\frac{\lambda v^2}{4m_\chi^2-m_h^2}\right)(2-3r_V+r_V^2)\right]\sqrt{1-r_V}\,.
\end{align}
where $N_c$ is the number of color for a fermion $f$, $r_x = m_x^2/m_{DM}^2$, $V=W^\pm,Z$ and $\delta_{W(Z)}=1(2)$. In the case where $A$ is the DM particle, the coupling $\lambda_{345}=\lambda_3+\lambda_4+\lambda_5$ becomes $\bar{\lambda}_{345} = \lambda_3 + \lambda_4 - \lambda_5$.

\subsection{Direct detection cross-sections} \label{sec:AppendB}

For the IS model, the DM-nucleon scattering cross-section is given by
\begin{equation}
	\sigma_{DM-N} = \frac{f_N^2(G^2+\tilde G^2)s_{2\theta}^2}{4\pi}\frac{m_{\rm DM}^2m_N^2}{(m_{\rm DM}+m_N)^2}\left(\frac{1}{m_\phi^2}-\frac{1}{m_h^2}\right)^2. \label{eq:ddxsec}
\end{equation}
where $f_N$ is the nucleon matrix element of the scalar current which is given by
\begin{equation}
	f_N \equiv \langle N|S_q|N\rangle = \frac29\frac{m_N}{v}\left(1+\frac72\sum_{q=u,d,s}f_{Tq}^{(N)}\right).
\end{equation}
For the $u$, $d$, and $s$ form factor, we use the following values~\cite{Bishara:2015cha}:
\begin{equation}
\begin{aligned}
	f^{(p)}_{Tu} &= (1.8\pm0.5)\times10^{-2},\\
	f^{(p)}_{Td} &= (3.4\pm1.1)\times10^{-2},\\
	f_{Ts}^{(p)} &= 0.043\pm0.011,
\end{aligned}
\qquad
\begin{aligned}
	f^{(n)}_{Tu} &= (1.6\pm0.5)\times10^{-2},\\
	f^{(n)}_{Td} &= (3.8\pm1.1)\times10^{-2},\\
	f_{Ts}^{(n)} &= 0.043\pm0.011.
\end{aligned}
\end{equation}

The DM-nucleon scattering cross-section for the ID model, in the case $H$ is the DM particle, is given by
\begin{equation}
    \sigma_{DM-N} = \frac{\lambda_{345}^2v^2}{m_h^4}\frac{f_N^2m_N^2}{(m_{DM}+m_N)^2}.
\end{equation}
Again, as in the case of DM annihilation, the DM-nucleon scattering cross-section in the case where $A$ is the DM particle can be obtained by a replacement $\lambda_{345}\to\bar{\lambda}_{345}$.\\

\section{The CTA sensitivity of other benchmarks}
\label{sec:AppendC}
\begin{figure}[!ht]
    \centering
    \begin{tabular}{cc}
        Draco dSph & Sculptor dSph \\
        \includegraphics[width=.4\textwidth]{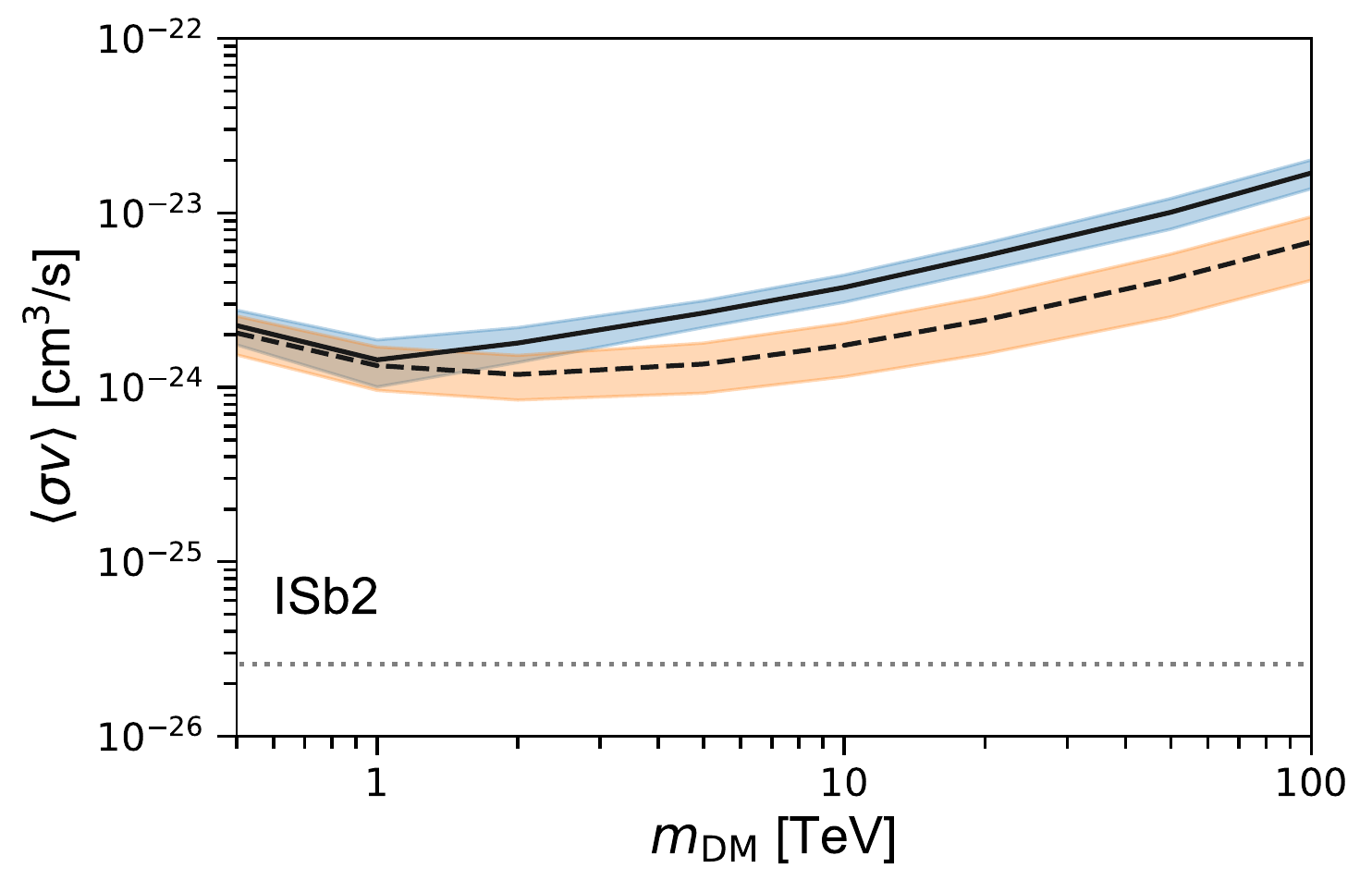} & \includegraphics[width=.4\textwidth]{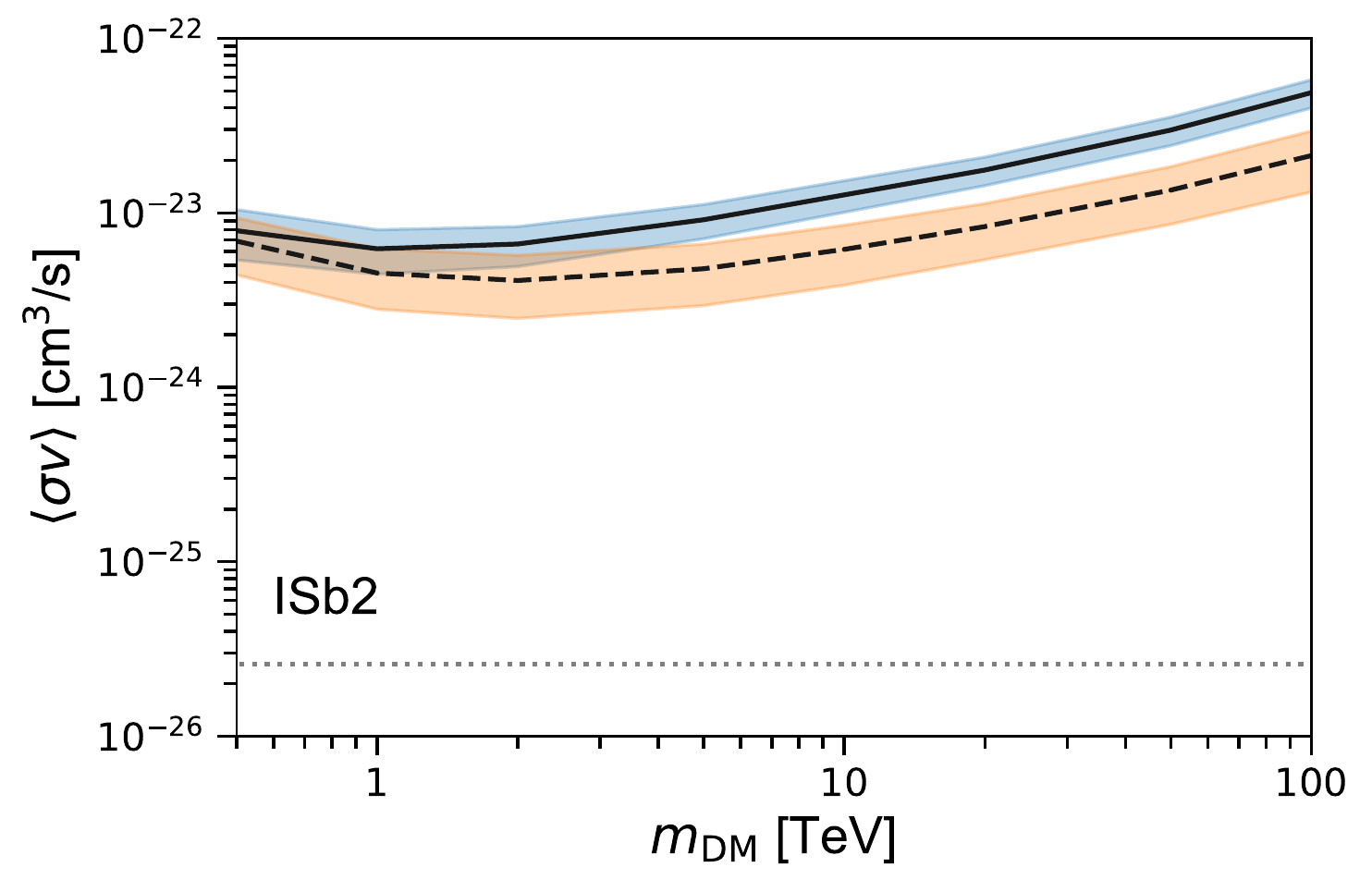}\\
        \includegraphics[width=.4\textwidth]{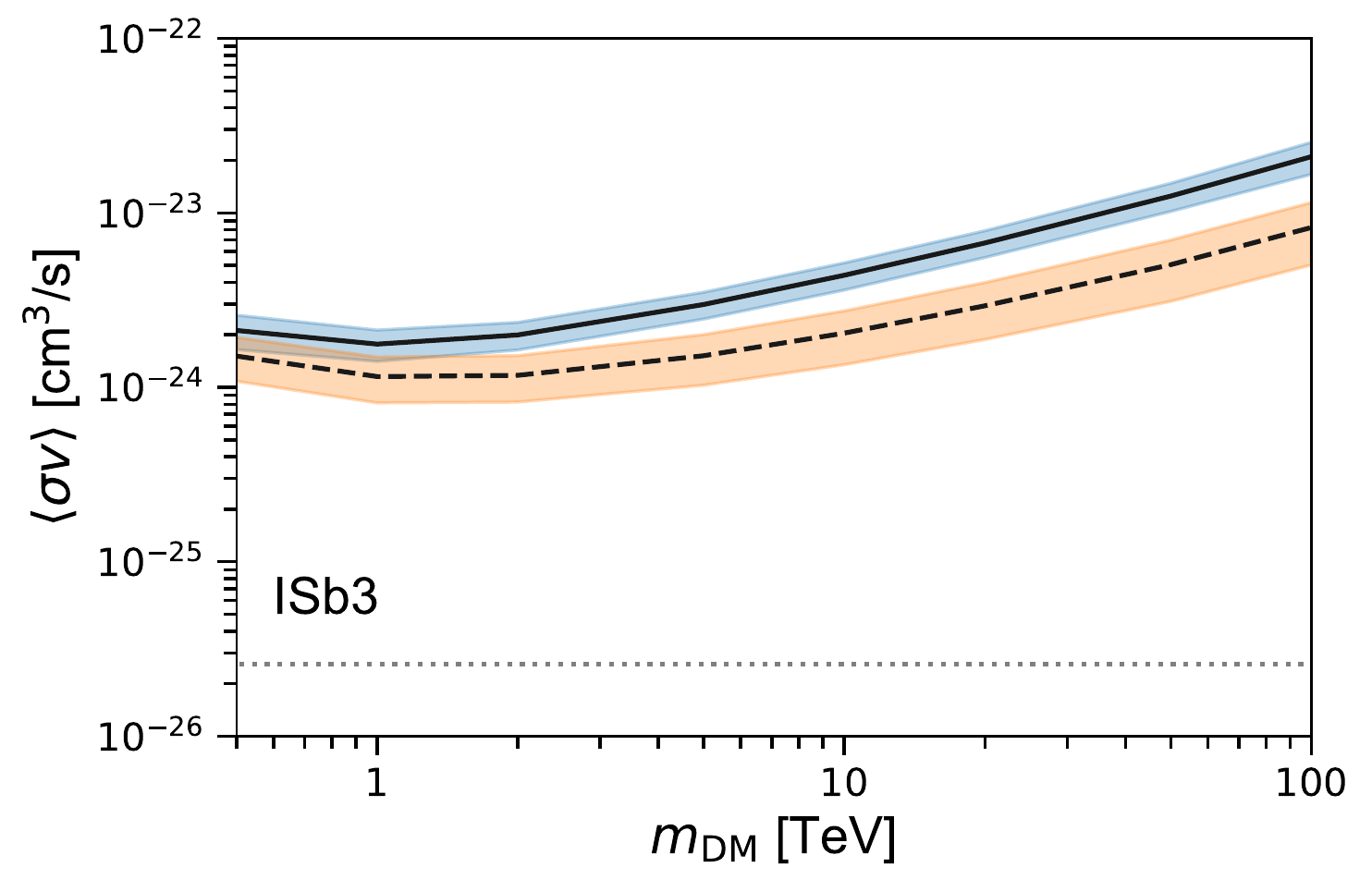} & \includegraphics[width=.4\textwidth]{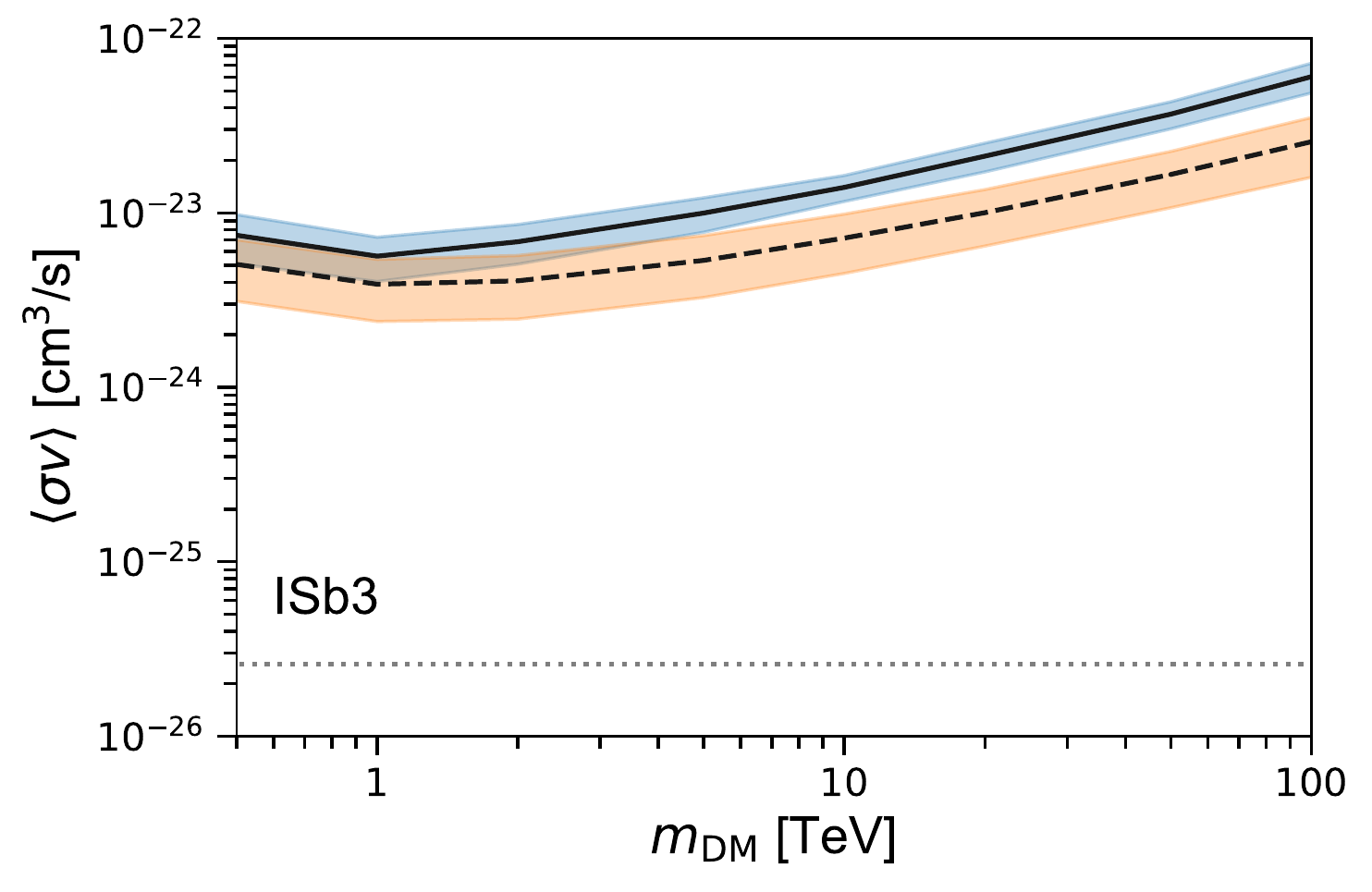}\\
        \includegraphics[width=.4\textwidth]{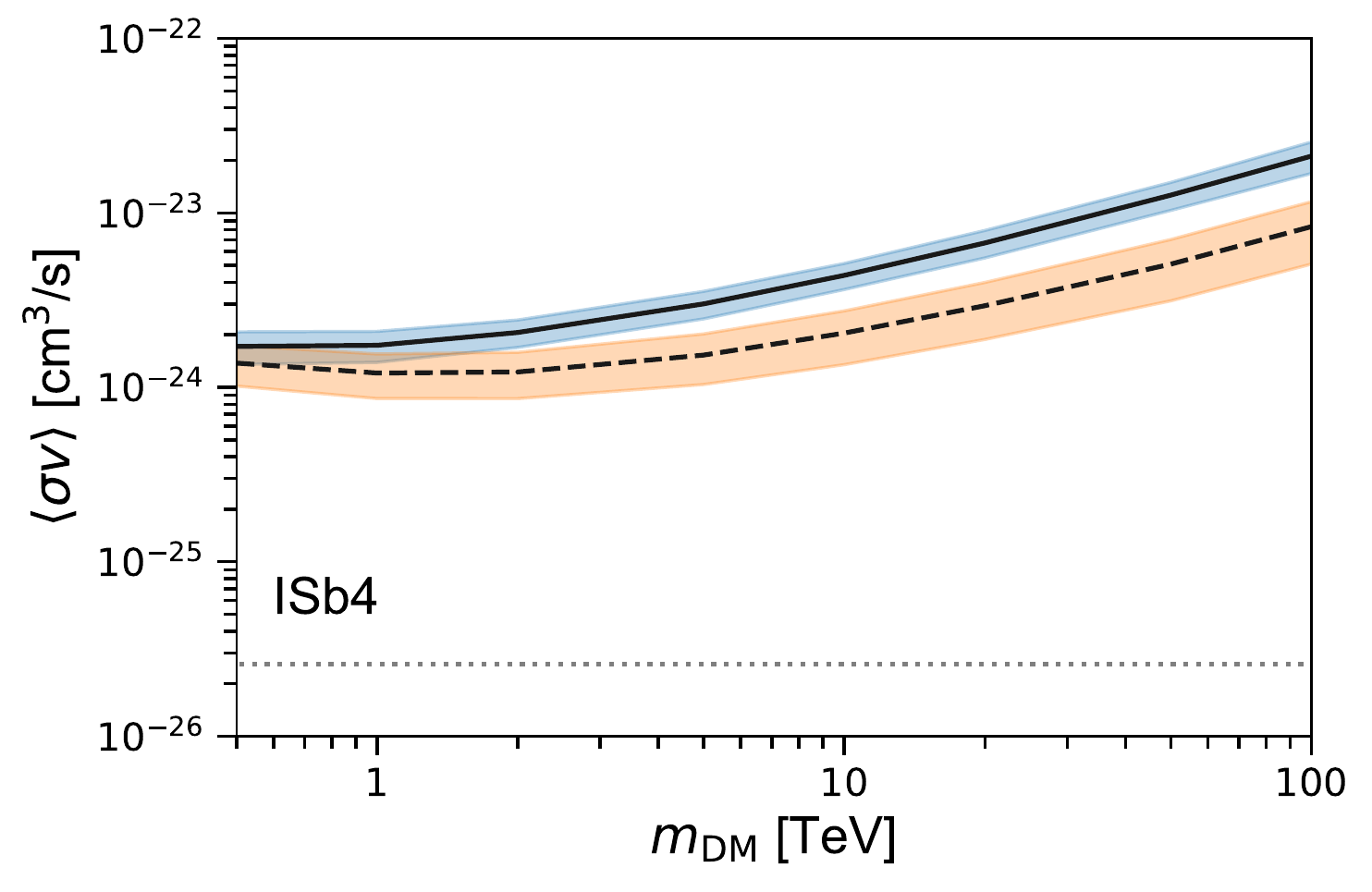} & \includegraphics[width=.4\textwidth]{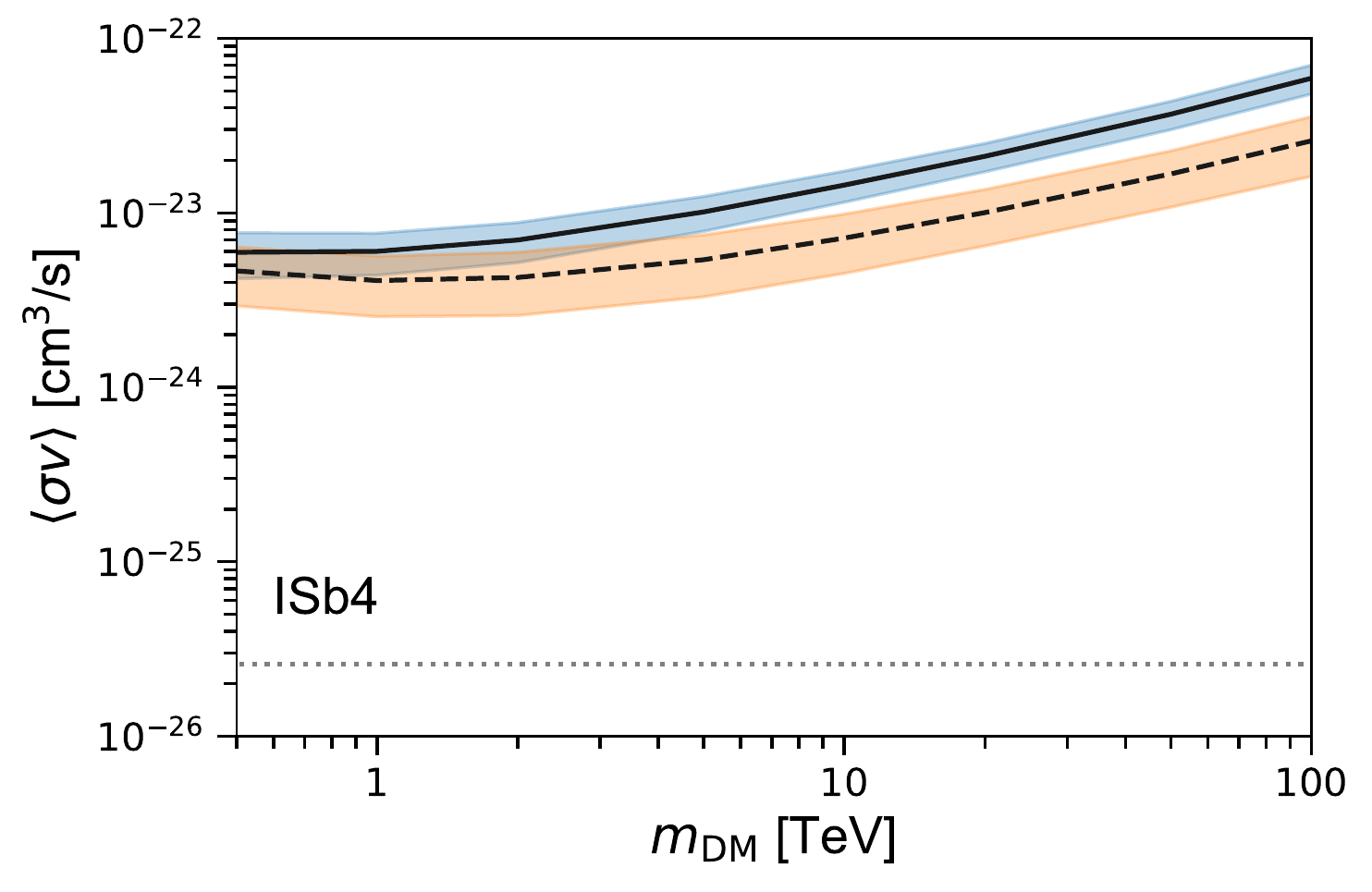}\\
    \end{tabular}
    \caption{CTA 5$\sigma$ discovery (solid) and 95\% CL exclusion (dashed) limits for the IS model benchmarks shown in Tab.~\ref{tab:benchmarkIS}. The shaded band represents the corresponding 1$\sigma$ uncertainty.  The dotted line represent the thermal cross-section.} 
    \label{fig:sensIS}
\end{figure}

\begin{figure}[!ht]
    \centering
    \begin{tabular}{cc}
        Draco dSph & Sculptor dSph \\
        \includegraphics[width=.4\textwidth]{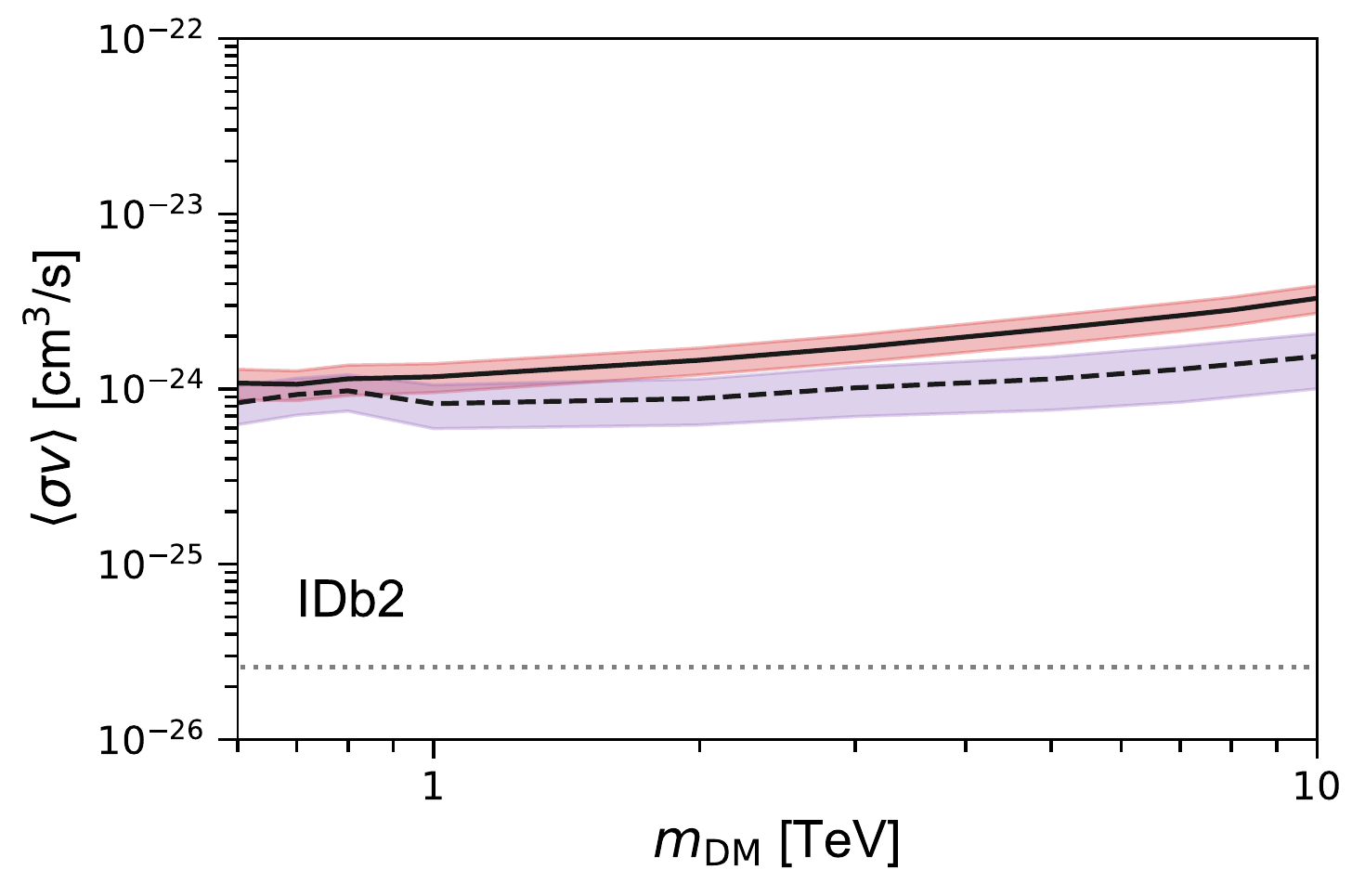} & \includegraphics[width=.4\textwidth]{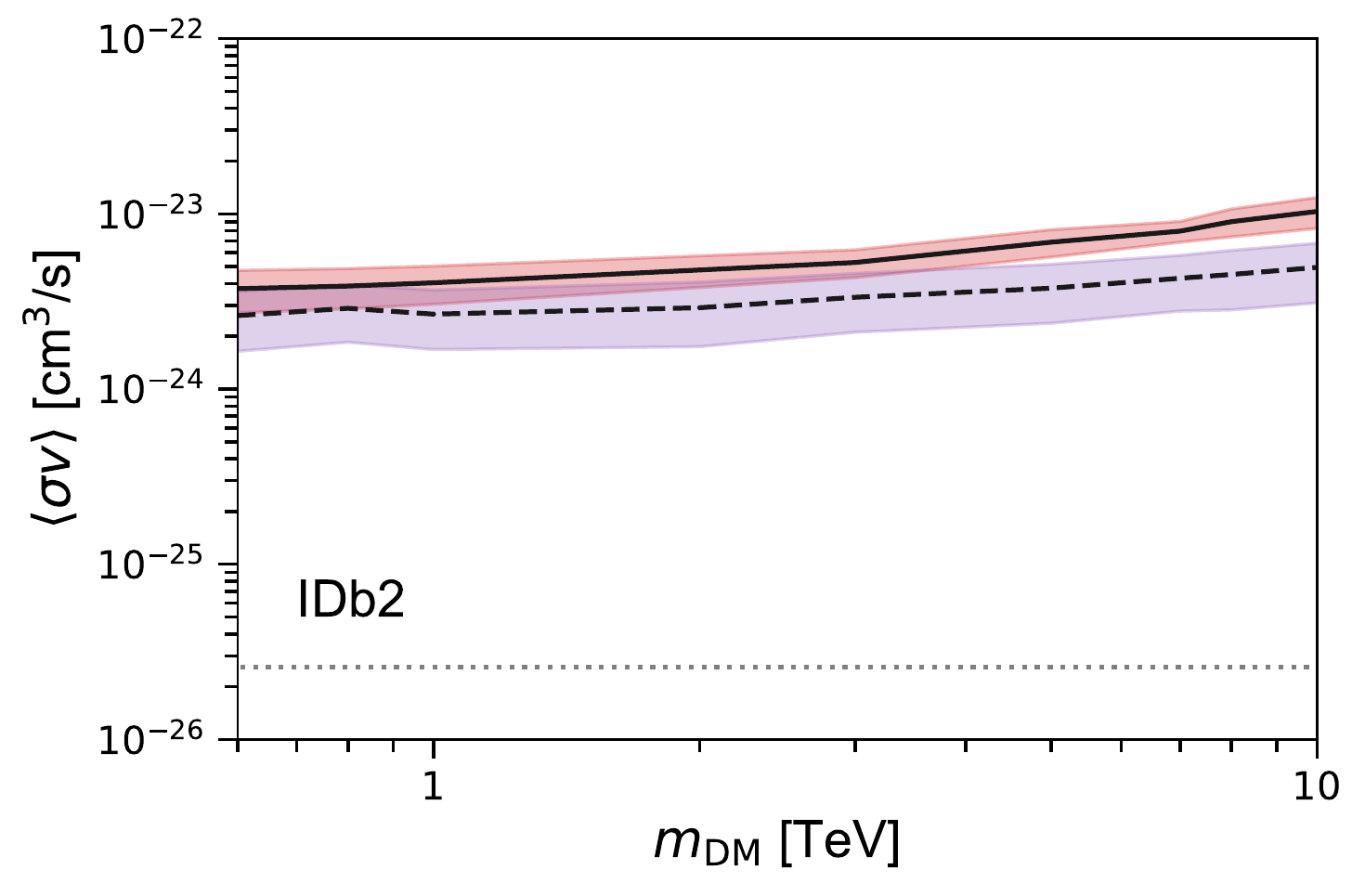}
    \end{tabular}
    \caption{CTA 5$\sigma$ discovery (solid) and 95\% CL exclusion (dashed) limits for the benchmark ID2. The shaded band represents the corresponding 1$\sigma$ uncertainty. The dotted line represent the thermal cross-section.}
    \label{fig:sensID}
\end{figure}

In this appendix we provide the CTA 5$\sigma$ detection and 95\% CL exclusion limit for the rest of the benchmark scenarios. For the IS model, the CTA limits for benchmarks ISb2-4, see Tab.~\ref{tab:benchmarkIS}, are shown in Fig.~\ref{fig:sensIS}. The CTA sensitivity for the benchmark IDb2 is shown in Fig.~\ref{fig:sensID}.

\newpage

\bibliographystyle{JHEP}
\bibliography{biblio.bib}

\providecommand{\href}[2]{#2}\begingroup\raggedright\begin{thebibliography}{10}

\bibitem{Planck_Cosmological_parameters}
{Planck Collaboration}, P.A.R.~{Ade}, N.~{Aghanim}, M.~{Arnaud}, M.~{Ashdown},
  J.~{Aumont} et~al., \emph{{Planck 2015 results. XIII. Cosmological
  parameters}},
  \href{https://doi.org/10.1051/0004-6361/201525830}{\emph{Astron. Astrophys}
  {\bfseries 594} (2016) A13}
  [\href{https://arxiv.org/abs/1502.01589}{{\ttfamily 1502.01589}}].

\bibitem{Bertone_etal_Dark_Matter_Review}
G.~{Bertone} and D.~{Hooper}, \emph{{History of dark matter}},
  \href{https://doi.org/10.1103/RevModPhys.90.045002}{\emph{Rev. Mod. Phys.}
  {\bfseries 90} (2018) 045002}
  [\href{https://arxiv.org/abs/1605.04909}{{\ttfamily 1605.04909}}].

\bibitem{Jungman1996SupersymmetricDM}
G.~{Jungman}, M.~{Kamionkowski} and K.~{Griest}, \emph{{Supersymmetric dark
  matter}}, \href{https://doi.org/10.1016/0370-1573(95)00058-5}{\emph{Phys.
  Rep.} {\bfseries 267} (1996) 195}
  [\href{https://arxiv.org/abs/hep-ph/9506380}{{\ttfamily hep-ph/9506380}}].

\bibitem{Bertone2005ParticleDM}
G.~{Bertone}, D.~{Hooper} and J.~{Silk}, \emph{{Particle dark matter: evidence,
  candidates and constraints}},
  \href{https://doi.org/10.1016/j.physrep.2004.08.031}{\emph{Phys. Rep.}
  {\bfseries 405} (2005) 279}
  [\href{https://arxiv.org/abs/hep-ph/0404175}{{\ttfamily hep-ph/0404175}}].

\bibitem{Feng2010DMCandidates}
J.L.~{Feng}, \emph{{Dark Matter Candidates from Particle Physics and Methods of
  Detection}},
  \href{https://doi.org/10.1146/annurev-astro-082708-101659}{\emph{Annu. Rev.
  Astron. Astrophys.} {\bfseries 48} (2010) 495}
  [\href{https://arxiv.org/abs/1003.0904}{{\ttfamily 1003.0904}}].

\bibitem{Arcadi2018WIMPs}
G.~{Arcadi}, M.~{Dutra}, P.~{Ghosh}, M.~{Lindner}, Y.~{Mambrini}, M.~{Pierre}
  et~al., \emph{{The waning of the WIMP? A review of models, searches, and
  constraints}},
  \href{https://doi.org/10.1140/epjc/s10052-018-5662-y}{\emph{Eur. Phys. J. C}
  {\bfseries 78} (2018) 203}
  [\href{https://arxiv.org/abs/1703.07364}{{\ttfamily 1703.07364}}].

\bibitem{Bertone2018DMNewEra}
G.~{Bertone} and T.M.P.~{Tait}, \emph{{A new era in the search for dark
  matter}}, \href{https://doi.org/10.1038/s41586-018-0542-z}{\emph{Nature}
  {\bfseries 562} (2018) 51}
  [\href{https://arxiv.org/abs/1810.01668}{{\ttfamily 1810.01668}}].

\bibitem{EuropeanStrategyforParticlePhysicsPreparatoryGroup:2019qin}
R.K.~Ellis et~al., \emph{{Physics Briefing Book}: {Input for the European
  Strategy for Particle Physics Update 2020}},
  \href{https://arxiv.org/abs/1910.11775}{{\ttfamily 1910.11775}}.

\bibitem{PAMELA:2017bna}
{\scshape PAMELA} collaboration, \emph{{Ten years of PAMELA in space}},
  \href{https://doi.org/10.1393/ncr/i2017-10140-x}{\emph{Riv. Nuovo Cim.}
  {\bfseries 40} (2017) 473}
  [\href{https://arxiv.org/abs/1801.10310}{{\ttfamily 1801.10310}}].

\bibitem{AMS:2021nhj}
{\scshape AMS} collaboration, \emph{{The Alpha Magnetic Spectrometer (AMS) on
  the international space station: Part II \textemdash{} Results from the first
  seven years}},
  \href{https://doi.org/10.1016/j.physrep.2020.09.003}{\emph{Phys. Rept.}
  {\bfseries 894} (2021) 1}.

\bibitem{Baur:2019jwm}
{\scshape IceCube} collaboration, \emph{{Dark matter searches with the IceCube
  Upgrade}}, \href{https://doi.org/10.22323/1.358.0506}{\emph{PoS} {\bfseries
  ICRC2019} (2020) 506} [\href{https://arxiv.org/abs/1908.08236}{{\ttfamily
  1908.08236}}].

\bibitem{Fermi-LAT:2013sme}
{\scshape Fermi-LAT} collaboration, \emph{{Dark Matter Constraints from
  Observations of 25 Milky Way Satellite Galaxies with the Fermi Large Area
  Telescope}}, \href{https://doi.org/10.1103/PhysRevD.89.042001}{\emph{Phys.
  Rev. D} {\bfseries 89} (2014) 042001}
  [\href{https://arxiv.org/abs/1310.0828}{{\ttfamily 1310.0828}}].

\bibitem{HESS_DMsearch_WLM}
H.~{Abdallah}, R.~{Adam}, F.~{Aharonian}, F.A.~{Benkhali}, E.O.~{Ang{\"u}ner},
  C.~{Arcaro} et~al., \emph{{Search for dark matter annihilation in the
  Wolf-Lundmark-Melotte dwarf irregular galaxy with H.E.S.S.}},
  \href{https://doi.org/10.1103/PhysRevD.103.102002}{\emph{Physical Review D}
  {\bfseries 103} (2021) 102002}
  [\href{https://arxiv.org/abs/2105.04325}{{\ttfamily 2105.04325}}].

\bibitem{HESS_DMsearch_DES}
H.~{Abdallah}, R.~{Adam}, F.~{Aharonian}, F.~{Ait Benkhali},
  E.O.~{Ang{\"u}ner}, M.~{Arakawa} et~al., \emph{{Search for dark matter
  signals towards a selection of recently detected DES dwarf galaxy satellites
  of the Milky Way with H.E.S.S.}},
  \href{https://doi.org/10.1103/PhysRevD.102.062001}{\emph{Physical Review D}
  {\bfseries 102} (2020) 062001}
  [\href{https://arxiv.org/abs/2008.00688}{{\ttfamily 2008.00688}}].

\bibitem{MAGIC_DMsearch_Triangulum}
V.A.~{Acciari}, S.~{Ansoldi}, L.A.~{Antonelli}, A.~{Arbet Engels}, D.~{Baack},
  A.~{Babi{\'c}} et~al., \emph{{A search for dark matter in Triangulum II with
  the MAGIC telescopes}},
  \href{https://doi.org/10.1016/j.dark.2020.100529}{\emph{Physics of the Dark
  Universe} {\bfseries 28} (2020) 100529}
  [\href{https://arxiv.org/abs/2003.05260}{{\ttfamily 2003.05260}}].

\bibitem{MAGIC_DMsearch_UrsaMajor}
M.L.~{Ahnen}, S.~{Ansoldi}, L.A.~{Antonelli}, C.~{Arcaro}, D.~{Baack},
  A.~{Babi{\'c}} et~al., \emph{{Indirect dark matter searches in the dwarf
  satellite galaxy Ursa Major II with the MAGIC telescopes}},
  \href{https://doi.org/10.1088/1475-7516/2018/03/009}{\emph{Journal of
  Cosmology and Astroparticle Physics} {\bfseries 2018} (2018) 009}
  [\href{https://arxiv.org/abs/1712.03095}{{\ttfamily 1712.03095}}].

\bibitem{VERITAS_DMsearch_dsph}
S.~{Archambault}, A.~{Archer}, W.~{Benbow}, R.~{Bird}, E.~{Bourbeau},
  T.~{Brantseg} et~al., \emph{{Dark matter constraints from a joint analysis of
  dwarf Spheroidal galaxy observations with VERITAS}},
  \href{https://doi.org/10.1103/PhysRevD.95.082001}{\emph{Physical Review D}
  {\bfseries 95} (2017) 082001}
  [\href{https://arxiv.org/abs/1703.04937}{{\ttfamily 1703.04937}}].

\bibitem{Winter2016}
M.~{Winter}, G.~{Zaharijas}, K.~{Bechtol} and J.~{Vandenbroucke},
  \emph{{Estimating the GeV Emission of Millisecond Pulsars in Dwarf Spheroidal
  Galaxies}},
  \href{https://doi.org/10.3847/2041-8205/832/1/L6}{\emph{Astrophys. J. Lett.}
  {\bfseries 832} (2016) L6}
  [\href{https://arxiv.org/abs/1607.06390}{{\ttfamily 1607.06390}}].

\bibitem{CTAConcept2013}
B.S.~{Acharya}, M.~{Actis}, T.~{Aghajani}, G.~{Agnetta}, J.~{Aguilar},
  F.~{Aharonian} et~al., \emph{{Introducing the CTA concept}},
  \href{https://doi.org/10.1016/j.astropartphys.2013.01.007}{\emph{Astropart.
  Phys.} {\bfseries 43} (2013) 3}.

\bibitem{CTAbook}
{\scshape CTA Consortium} collaboration, B.S.~Acharya et~al., \emph{{Science
  with the Cherenkov Telescope Array}}, WSP (11, 2018),
  \href{https://doi.org/10.1142/10986}{10.1142/10986},
  [\href{https://arxiv.org/abs/1709.07997}{{\ttfamily 1709.07997}}].

\bibitem{Profumo:2010kp}
S.~Profumo, L.~Ubaldi and C.~Wainwright, \emph{{Singlet Scalar Dark Matter:
  monochromatic gamma rays and metastable vacua}},
  \href{https://doi.org/10.1103/PhysRevD.82.123514}{\emph{Phys. Rev. D}
  {\bfseries 82} (2010) 123514}
  [\href{https://arxiv.org/abs/1009.5377}{{\ttfamily 1009.5377}}].

\bibitem{Buckley:2014fba}
M.R.~Buckley, D.~Feld and D.~Goncalves, \emph{{Scalar Simplified Models for
  Dark Matter}}, \href{https://doi.org/10.1103/PhysRevD.91.015017}{\emph{Phys.
  Rev. D} {\bfseries 91} (2015) 015017}
  [\href{https://arxiv.org/abs/1410.6497}{{\ttfamily 1410.6497}}].

\bibitem{Abdallah:2014hon}
J.~Abdallah et~al., \emph{{Simplified Models for Dark Matter and Missing Energy
  Searches at the LHC}},  \href{https://arxiv.org/abs/1409.2893}{{\ttfamily
  1409.2893}}.

\bibitem{Feng:2014vea}
L.~Feng, S.~Profumo and L.~Ubaldi, \emph{{Closing in on singlet scalar dark
  matter: LUX, invisible Higgs decays and gamma-ray lines}},
  \href{https://doi.org/10.1007/JHEP03(2015)045}{\emph{JHEP} {\bfseries 03}
  (2015) 045} [\href{https://arxiv.org/abs/1412.1105}{{\ttfamily 1412.1105}}].

\bibitem{Bishara:2015cha}
F.~Bishara, J.~Brod, P.~Uttayarat and J.~Zupan, \emph{{Nonstandard Yukawa
  Couplings and Higgs Portal Dark Matter}},
  \href{https://doi.org/10.1007/JHEP01(2016)010}{\emph{JHEP} {\bfseries 01}
  (2016) 010} [\href{https://arxiv.org/abs/1504.04022}{{\ttfamily
  1504.04022}}].

\bibitem{Altmannshofer:2019wjb}
W.~Altmannshofer, B.~Maddock and S.~Profumo, \emph{{Doubly Blind Spots in
  Scalar Dark Matter Models}},
  \href{https://doi.org/10.1103/PhysRevD.100.055033}{\emph{Phys. Rev. D}
  {\bfseries 100} (2019) 055033}
  [\href{https://arxiv.org/abs/1907.01726}{{\ttfamily 1907.01726}}].

\bibitem{Pongkitivanichkul:2019cvm}
C.~Pongkitivanichkul, N.~Thongyoi and P.~Uttayarat, \emph{{Inverse seesaw
  mechanism and portal dark matter}},
  \href{https://doi.org/10.1103/PhysRevD.100.035034}{\emph{Phys. Rev. D}
  {\bfseries 100} (2019) 035034}
  [\href{https://arxiv.org/abs/1905.13224}{{\ttfamily 1905.13224}}].

\bibitem{Deshpande:1977rw}
N.G.~Deshpande and E.~Ma, \emph{{Pattern of Symmetry Breaking with Two Higgs
  Doublets}}, \href{https://doi.org/10.1103/PhysRevD.18.2574}{\emph{Phys. Rev.
  D} {\bfseries 18} (1978) 2574}.

\bibitem{LopezHonorez:2006gr}
L.~Lopez~Honorez, E.~Nezri, J.F.~Oliver and M.H.G.~Tytgat, \emph{{The Inert
  Doublet Model: An Archetype for Dark Matter}},
  \href{https://doi.org/10.1088/1475-7516/2007/02/028}{\emph{JCAP} {\bfseries
  02} (2007) 028} [\href{https://arxiv.org/abs/hep-ph/0612275}{{\ttfamily
  hep-ph/0612275}}].

\bibitem{Goudelis:2013uca}
A.~Goudelis, B.~Herrmann and O.~St\r{a}l, \emph{{Dark matter in the Inert
  Doublet Model after the discovery of a Higgs-like boson at the LHC}},
  \href{https://doi.org/10.1007/JHEP09(2013)106}{\emph{JHEP} {\bfseries 09}
  (2013) 106} [\href{https://arxiv.org/abs/1303.3010}{{\ttfamily 1303.3010}}].

\bibitem{Arcadi:2019lka}
G.~Arcadi, A.~Djouadi and M.~Raidal, \emph{{Dark Matter through the Higgs
  portal}}, \href{https://doi.org/10.1016/j.physrep.2019.11.003}{\emph{Phys.
  Rept.} {\bfseries 842} (2020) 1}
  [\href{https://arxiv.org/abs/1903.03616}{{\ttfamily 1903.03616}}].

\bibitem{XENON:2020kmp}
{\scshape XENON} collaboration, \emph{{Projected WIMP sensitivity of the
  XENONnT dark matter experiment}},
  \href{https://doi.org/10.1088/1475-7516/2020/11/031}{\emph{JCAP} {\bfseries
  11} (2020) 031} [\href{https://arxiv.org/abs/2007.08796}{{\ttfamily
  2007.08796}}].

\bibitem{Khachatryan:2016vau}
{\scshape ATLAS, CMS} collaboration, \emph{{Measurements of the Higgs boson
  production and decay rates and constraints on its couplings from a combined
  ATLAS and CMS analysis of the LHC pp collision data at $ \sqrt{s}=7 $ and 8
  TeV}}, \href{https://doi.org/10.1007/JHEP08(2016)045}{\emph{JHEP} {\bfseries
  08} (2016) 045} [\href{https://arxiv.org/abs/1606.02266}{{\ttfamily
  1606.02266}}].

\bibitem{CMS:2018uag}
{\scshape CMS} collaboration, \emph{{Combined measurements of Higgs boson
  couplings in proton\textendash{}proton collisions at $\sqrt{s}=13\,\text
  {Te}\text {V} $}},
  \href{https://doi.org/10.1140/epjc/s10052-019-6909-y}{\emph{Eur. Phys. J. C}
  {\bfseries 79} (2019) 421}
  [\href{https://arxiv.org/abs/1809.10733}{{\ttfamily 1809.10733}}].

\bibitem{ATLAS:2019nkf}
{\scshape ATLAS} collaboration, \emph{{Combined measurements of Higgs boson
  production and decay using up to $80$ fb$^{-1}$ of proton-proton collision
  data at $\sqrt{s}=$ 13 TeV collected with the ATLAS experiment}},
  \href{https://doi.org/10.1103/PhysRevD.101.012002}{\emph{Phys. Rev. D}
  {\bfseries 101} (2020) 012002}
  [\href{https://arxiv.org/abs/1909.02845}{{\ttfamily 1909.02845}}].

\bibitem{Treesukrat:2019ahh}
W.~Treesukrat and P.~Uttayarat, \emph{{Dark matter from the inert Higgs doublet
  model}}, \href{https://doi.org/10.1088/1742-6596/1380/1/012093}{\emph{J.
  Phys. Conf. Ser.} {\bfseries 1380} (2019) 012093}.

\bibitem{Ivanov:2006yq}
I.P.~Ivanov, \emph{Minkowski space structure of the higgs potential in 2hdm},
  \href{https://doi.org/10.1103/PhysRevD.76.039902,
  10.1103/PhysRevD.75.035001}{\emph{Phys. Rev. D} {\bfseries 75} (2007)
  035001}.

\bibitem{Ginzburg:2003fe}
I.F.~Ginzburg and I.P.~Ivanov, \emph{{Tree level unitarity constraints in the
  2HDM with CP violation}},
  \href{https://arxiv.org/abs/hep-ph/0312374}{{\ttfamily hep-ph/0312374}}.

\bibitem{ctools}
J.~{Kn{\"o}dlseder}, M.~{Mayer}, C.~{Deil}, J.B.~{Cayrou}, E.~{Owen},
  N.~{Kelley-Hoskins} et~al., \emph{{GammaLib and ctools. A software framework
  for the analysis of astronomical gamma-ray data}},
  \href{https://doi.org/10.1051/0004-6361/201628822}{\emph{Astronomy and
  Astrophysics} {\bfseries 593} (2016) A1}
  [\href{https://arxiv.org/abs/1606.00393}{{\ttfamily 1606.00393}}].

\bibitem{irf-prod3}
C.T.A.~Observatory and C.T.A.~Consortium, \emph{{CTAO Instrument Response
  Functions - version prod3b-v2}},
  \href{https://doi.org/10.5281/zenodo.5163273}{\emph{Zenodo} (2016) }.

\bibitem{2015DMinDsph}
V.~Bonnivard, C.~Combet, M.~Daniel, S.~Funk, A.~Geringer-Sameth, J.A.~Hinton
  et~al., \emph{Dark matter annihilation and decay in dwarf spheroidal
  galaxies: the classical and ultrafaint dsphs},
  \href{https://doi.org/10.1093/mnras/stv1601}{\emph{Monthly Notices of the
  Royal Astronomical Society} {\bfseries 453} (2015) 849–867}.

\bibitem{Aguirre_Santaella_2020}
A.~Aguirre-Santaella, V.~Gammaldi, M.~Sánchez-Conde and D.~Nieto,
  \emph{Cherenkov telescope array sensitivity to branon dark matter models},
  \href{https://doi.org/10.1088/1475-7516/2020/10/041}{\emph{Journal of
  Cosmology and Astroparticle Physics} {\bfseries 2020} (2020) 041–041}.

\bibitem{Griest:1990kh}
K.~Griest and D.~Seckel, \emph{{Three exceptions in the calculation of relic
  abundances}}, \href{https://doi.org/10.1103/PhysRevD.43.3191}{\emph{Phys.
  Rev. D} {\bfseries 43} (1991) 3191}.

\bibitem{DarkSide-20k:2017zyg}
{\scshape DarkSide-20k} collaboration, \emph{{DarkSide-20k: A 20 tonne
  two-phase LAr TPC for direct dark matter detection at LNGS}},
  \href{https://doi.org/10.1140/epjp/i2018-11973-4}{\emph{Eur. Phys. J. Plus}
  {\bfseries 133} (2018) 131}
  [\href{https://arxiv.org/abs/1707.08145}{{\ttfamily 1707.08145}}].

\bibitem{LUX-ZEPLIN:2018poe}
{\scshape LUX-ZEPLIN} collaboration, \emph{{Projected WIMP sensitivity of the
  LUX-ZEPLIN dark matter experiment}},
  \href{https://doi.org/10.1103/PhysRevD.101.052002}{\emph{Phys. Rev. D}
  {\bfseries 101} (2020) 052002}
  [\href{https://arxiv.org/abs/1802.06039}{{\ttfamily 1802.06039}}].

\bibitem{PandaX:2018wtu}
{\scshape PandaX} collaboration, \emph{{Dark matter direct search sensitivity
  of the PandaX-4T experiment}},
  \href{https://doi.org/10.1007/s11433-018-9259-0}{\emph{Sci. China Phys. Mech.
  Astron.} {\bfseries 62} (2019) 31011}
  [\href{https://arxiv.org/abs/1806.02229}{{\ttfamily 1806.02229}}].

\bibitem{IceCube:2014rqf}
{\scshape IceCube} collaboration, \emph{{Multipole analysis of IceCube data to
  search for dark matter accumulated in the Galactic halo}},
  \href{https://doi.org/10.1140/epjc/s10052-014-3224-5}{\emph{Eur. Phys. J. C}
  {\bfseries 75} (2015) 20} [\href{https://arxiv.org/abs/1406.6868}{{\ttfamily
  1406.6868}}].

\bibitem{Dsph_DM_search_MAGIC}
{MAGIC Collaboration}, V.A.~{Acciari}, S.~{Ansoldi}, L.A.~{Antonelli},
  A.~{Arbet Engels}, M.~{Artero} et~al., \emph{{Combined searches for dark
  matter in dwarf spheroidal galaxies observed with the MAGIC telescopes,
  including new data from Coma Berenices and Draco}}, {\emph{arXiv e-prints}
  (2021) arXiv:2111.15009} [\href{https://arxiv.org/abs/2111.15009}{{\ttfamily
  2111.15009}}].

\bibitem{IceCube_draco_UL}
M.G.~{Aartsen}, R.~{Abbasi}, Y.~{Abdou}, M.~{Ackermann}, J.~{Adams},
  J.A.~{Aguilar} et~al., \emph{{IceCube search for dark matter annihilation in
  nearby galaxies and galaxy clusters}},
  \href{https://doi.org/10.1103/PhysRevD.88.122001}{\emph{Physical Review D}
  {\bfseries 88} (2013) 122001}
  [\href{https://arxiv.org/abs/1307.3473}{{\ttfamily 1307.3473}}].

\bibitem{LHAASO_dsph_UL}
D.-Z.~{He}, X.-J.~{Bi}, S.-J.~{Lin}, P.-F.~{Yin} and X.~{Zhang},
  \emph{{Prospect for dark matter signatures from dwarf galaxies by LHAASO}},
  {\emph{arXiv e-prints} (2019) arXiv:1903.11910}
  [\href{https://arxiv.org/abs/1903.11910}{{\ttfamily 1903.11910}}].

\bibitem{Gammaldi:2016uhg}
V.~Gammaldi, V.~Avila-Reese, O.~Valenzuela and A.X.~Gonzales-Morales,
  \emph{{Analysis of the very inner Milky Way dark matter distribution and
  gamma-ray signals}},
  \href{https://doi.org/10.1103/PhysRevD.94.121301}{\emph{Phys. Rev. D}
  {\bfseries 94} (2016) 121301}
  [\href{https://arxiv.org/abs/1607.02012}{{\ttfamily 1607.02012}}].

\bibitem{Gondolo:1999ef}
P.~Gondolo and J.~Silk, \emph{{Dark matter annihilation at the galactic
  center}}, \href{https://doi.org/10.1103/PhysRevLett.83.1719}{\emph{Phys. Rev.
  Lett.} {\bfseries 83} (1999) 1719}
  [\href{https://arxiv.org/abs/astro-ph/9906391}{{\ttfamily
  astro-ph/9906391}}].

\bibitem{Sanchez-Conde:2013yxa}
M.A.~S\'anchez-Conde and F.~Prada, \emph{{The flattening of the
  concentration\textendash{}mass relation towards low halo masses and its
  implications for the annihilation signal boost}},
  \href{https://doi.org/10.1093/mnras/stu1014}{\emph{Mon. Not. Roy. Astron.
  Soc.} {\bfseries 442} (2014) 2271}
  [\href{https://arxiv.org/abs/1312.1729}{{\ttfamily 1312.1729}}].

\bibitem{Moline:2016pbm}
A.~Molin\'e, M.A.~S\'anchez-Conde, S.~Palomares-Ruiz and F.~Prada,
  \emph{{Characterization of subhalo structural properties and implications for
  dark matter annihilation signals}},
  \href{https://doi.org/10.1093/mnras/stx026}{\emph{Mon. Not. Roy. Astron.
  Soc.} {\bfseries 466} (2017) 4974}
  [\href{https://arxiv.org/abs/1603.04057}{{\ttfamily 1603.04057}}].

\bibitem{Ando:2019xlm}
S.~Ando, T.~Ishiyama and N.~Hiroshima, \emph{{Halo Substructure Boosts to the
  Signatures of Dark Matter Annihilation}},
  \href{https://doi.org/10.3390/galaxies7030068}{\emph{Galaxies} {\bfseries 7}
  (2019) 68} [\href{https://arxiv.org/abs/1903.11427}{{\ttfamily 1903.11427}}].

\bibitem{Hisano:2004ds}
J.~Hisano, S.~Matsumoto, M.M.~Nojiri and O.~Saito, \emph{{Non-perturbative
  effect on dark matter annihilation and gamma ray signature from galactic
  center}}, \href{https://doi.org/10.1103/PhysRevD.71.063528}{\emph{Phys. Rev.
  D} {\bfseries 71} (2005) 063528}
  [\href{https://arxiv.org/abs/hep-ph/0412403}{{\ttfamily hep-ph/0412403}}].

\bibitem{Lattanzi:2008qa}
M.~Lattanzi and J.I.~Silk, \emph{{Can the WIMP annihilation boost factor be
  boosted by the Sommerfeld enhancement?}},
  \href{https://doi.org/10.1103/PhysRevD.79.083523}{\emph{Phys. Rev. D}
  {\bfseries 79} (2009) 083523}
  [\href{https://arxiv.org/abs/0812.0360}{{\ttfamily 0812.0360}}].

\bibitem{Combined_DM_searches}
C.~{Armand}, E.~{Charles}, M.~{di Mauro}, C.~{Giuri}, J.P.~{Harding},
  D.~{Kerszberg} et~al., \emph{{Combined dark matter searches towards dwarf
  spheroidal galaxies with Fermi-LAT, HAWC, H.E.S.S., MAGIC, and VERITAS}},
  {\emph{arXiv e-prints} (2021) arXiv:2108.13646}
  [\href{https://arxiv.org/abs/2108.13646}{{\ttfamily 2108.13646}}].

\bibitem{Benito:2019ngh}
M.~Benito, A.~Cuoco and F.~Iocco, \emph{{Handling the Uncertainties in the
  Galactic Dark Matter Distribution for Particle Dark Matter Searches}},
  \href{https://doi.org/10.1088/1475-7516/2019/03/033}{\emph{JCAP} {\bfseries
  03} (2019) 033} [\href{https://arxiv.org/abs/1901.02460}{{\ttfamily
  1901.02460}}].

\bibitem{Benito:2020lgu}
M.~Benito, F.~Iocco and A.~Cuoco, \emph{{Uncertainties in the Galactic Dark
  Matter distribution: An update}},
  \href{https://doi.org/10.1016/j.dark.2021.100826}{\emph{Phys. Dark Univ.}
  {\bfseries 32} (2021) 100826}
  [\href{https://arxiv.org/abs/2009.13523}{{\ttfamily 2009.13523}}].

\bibitem{CTACollaboration2021}
A.~{Acharyya}, R.~{Adam}, C.~{Adams}, I.~{Agudo}, A.~{Aguirre-Santaella},
  R.~{Alfaro} et~al., \emph{{Sensitivity of the Cherenkov Telescope Array to a
  dark matter signal from the Galactic centre}},
  \href{https://doi.org/10.1088/1475-7516/2021/01/057}{\emph{J. Cosmol.
  Astropart. Phys.} {\bfseries 2021} (2021) 057}
  [\href{https://arxiv.org/abs/2007.16129}{{\ttfamily 2007.16129}}].

\end{thebibliography}\endgroup

\end{document}